\documentclass[aps,prb,twocolumn,superscriptaddress,longbibliography]{revtex4-1}%
\usepackage{graphicx}
\usepackage{float}
\usepackage{dcolumn}
\usepackage{bm,color}
\usepackage{amsmath,amssymb,dsfont,amstext,amsfonts}
\usepackage{extarrows}
\usepackage[colorlinks=true,linkcolor=blue,urlcolor=blue,citecolor=blue]{hyperref}
\usepackage{xcolor}
\usepackage{wasysym}
\usepackage{mathtools}
\usepackage{bbold}

\usepackage[normalem]{ulem} %makes underlining possible: \uline \uwave \sout
\usepackage{color}%makes \textcolor{grey}{text} available

\newcommand{\la}{\langle}
\newcommand{\ra}{\rangle}

\newcommand{\bq}{\bm{q}}

\def\bra#1{\mathinner{\langle{#1}|}}
\def\ket#1{\mathinner{|{#1}\rangle}}

\def\braket#1{\mathinner{\langle{#1}\rangle}}

\def\sgn{\mathrm{sgn}}

\def\tr{\mathrm{Tr}}
\def\dd{\mathrm{d}}
\def\id{\mathbb{1}} % need bbold,bm package

\def\bs#1{\boldsymbol{#1}}

\begin{document}
\title{ Nested-sphere description of the $N$-level Chern number and the generalized Bloch hypersphere}
\author{Cameron J.D. Kemp}
\affiliation{Robinson College, University of Cambridge, Grange Road, Cambridge  CB3 9AN, United Kingdom\looseness=-1}
\author{Nigel R. Cooper}
\affiliation{TCM Group, Cavendish Laboratory, University of Cambridge, JJ Thomson Avenue, Cambridge CB3 0HE, United Kingdom\looseness=-1}
\affiliation{Department of Physics and Astronomy, University of Florence, Via G. Sansone 1, 50019 Sesto Fiorentino, Italy\looseness=-1}
\author{F. Nur \"{U}nal}
\email{fnu20@cam.ac.uk}
\affiliation{TCM Group, Cavendish Laboratory, University of Cambridge, JJ Thomson Avenue, Cambridge CB3 0HE, United Kingdom\looseness=-1}

%TC:ignore
\date{\today}
\begin{abstract}
The geometric interpretation of (pseudo)spin $1/2$ systems on the Bloch sphere has been appreciated across different areas ranging from condensed matter to quantum information and high energy physics. Although similar notions for larger Hilbert spaces are established in mathematics, they have been so far less explored beyond the two-level case for practical usage in condensed matter settings, or have involved restrictions to sub manifolds within the full Hilbert space. 
We here employ a coherence vector description to theoretically characterize a general $N$-level system on the higher dimensional \textit{generalized} Bloch (hyper)sphere by respecting the structure of the underlying SU$(N)$ algebra and construct physically intuitive geometric pictures for topological concepts.
%We here theoretically characterize a general $N$-level system by its coherence vector described on the higher dimensional \textit{generalized} Bloch (hyper)sphere and construct physically intuitive geometric pictures for topological concepts that is rooted in the underlying SU$(N)$ algebra.
Focusing on two spatial dimensions, we reveal a geometric interpretation for the Chern number in larger Hilbert spaces in terms of a nested structure comprising $N-1$ two-spheres. We demonstrate that for the $N$-level case, there is an exterior two-sphere that provides a useful characterization of the system, notably by playing a primary role in determining the Chern number. The external sphere can be directly measured in ultracold atoms via well-established band mapping techniques, thereby imparting knowledge of the topological nature of state. We also investigate how the time evolution of the coherence vector defined on the generalized Bloch hypersphere can be utilized to extract the full state vector in experiments, allowing us to develop a tomography scheme involving quenches for three-level systems. Our geometric description opens up a new avenue for the interpretation of the topological classification and the dynamical illustration of multi-level systems, which in turn is anticipated to help in the design of new experimental probes.
%provides new understanding
%shines a new light
\end{abstract}
\maketitle
%TC:endignore

\section{Introduction}% 
As the minimal model for a vast range of phenomena, two-level quantum systems are repeatedly employed in physics research, where the Bloch sphere provides the paradigmatic visual aid for capturing the dynamics of interest. This geometric description is also useful in utilizing abstract concepts from mathematics to reveal novel topological properties in condensed matter systems~\cite{TKNN}. Most prominently, the Chern number of the simplest two-band models can be interpreted in terms of the number of wrappings of the Bloch sphere by the Hamiltonian vector~\cite{Xiao10_RMP,HasanKane10_RMP,QiZhang11_RMP}. This view extends to more complex out-of-equilibrium settings connecting to the links and knots the wave function ties on the two-sphere ($\text{S}^2$) during its dynamics, pertaining to underlying invariants of static or anomalous Floquet topological phases~\cite{Cooper19_RMP,Wang17_PRL_Chern,Tarnowski19_NatCom,Wintersperger20_NatPhys,Unal19_PRR,ChernHopf_Yu,SunWeiPan18_PRL_hopfExp,YiPan19_arx_hopfTori,ZHANG18_sciBul,HuZhao20_PRL}. The Bloch sphere picture even caters to exotic topologies with non-Abelian properties such as the Euler class~\cite{Unal20_PRL,Ahn19_PRX} which, although being essentially of multi-band nature, can be connected to a two-level description~\cite{BouhonSlagerBzdusek20_NatPhys_nonabelian,BouhonSlager20_PRB_FragileEuler,JiangSlager21_NatPhys_nonabelianExperiment}.

In addition to its eminent theoretical convenience, the Bloch sphere lays the foundation for various measurement protocols and constitutes a crucial description in certain experimental settings such as with superconducting qubits or well-controlled single spins~\cite{XiangNori13_RMP,Hanson07_RMP,Makhlin01_RMP}. While it remains widely elusive in condensed matter experiments for the observation of topologically protected phenomena in reciprocal space, where topology is usually detected via surface states or response functions such as the Hall conductivity~\cite{TKNN}, the versatility of quantum simulators has also made it possible to directly access the Bloch sphere in this context~\cite{Cooper19_RMP}. 
Indeed, sophisticated momentum- and time-resolved probing techniques developed in ultracold quantum gases allow for, e.g.,~the tomography of the full Bloch vector and reconstructing the Berry curvature ~\cite{Alba11_PRL,Hauke14_PRL,Flaschner16_Sci}, measuring the components of the Wilson line matrix~\cite{Li16_Science_wilson,WilczekZee84_PRL,AlexandradinataBernevig16_PRB,Alexandradinata14_PRB,BouhonSlager19_PRB}, 
accessing the $\mathds{Z}_2$ invariant~\cite{Grusdt14_PRA_Z2}, 
as well as the observation of topological singularities on the Bloch sphere~\cite{Tarnowski19_NatCom} further connecting also to dynamical phase transitions~\cite{Flaschner18_NatPhys_DynSingl}. %, as well as expanding our understanding of various topologically protected phenomena[]. 
%\note{photonic examples?}

Although similar hypersphere descriptions for higher dimensional Hilbert spaces exist theoretically~\cite{Hioe81_PRL,Kimura03_PLA,Graf21_arXiv_eigenprojectors,Khanna97_AnnalsPhys_GeometricPF,LeePark12_JPSocJapan}, they  naturally become less intuitive as the complexity increases. Considering the widespread implementation of the Bloch sphere in depicting two-level dynamics, there is an apparent disparity with the scarcity of such concrete geometric characterizations to take advantage of particularly for the study of topological properties in larger Hilbert space dimensions.  Certain definitions involving Euler angles have been also employed in superfluid helium 3 for the study of topology of defects that explore a sub manifold of the full three-level system~\cite{VollhardtWolfle90_He3book}, which hence remains restricted. Much as the topological classification for $N$-level systems is known, the question however persists whether a simple geometric picture in terms of a `sphere' can be found for the Chern number in an unrestricted form. This pursuit becomes more pressing upon considering that topology is closely linked to the underlying geometry of the states, underpinning connections to the quantum geometric tensor~\cite{Resta,Ahn21_arXiv_Riemann}, circular dichroism~\cite{Asteria19_NatPhys,TanYu19_PRL_QGTmeasurement,Tran17_SciAdv_dichrosim,TranCooperGoldman18_PRA_circdichroLL} and quantum metric~\cite{OzawaGoldman18_PRB_metric,OzawaMera21_PRB,MeraZhangGoldman22_SciPostPhys,Zhang_2022}, which all have found extensive application in experiments in recent years. Establishing an explicit geometric interpretation for $N$-level systems would not only provide novel intuitive insight,
%~\cite{Kimura03_PLA,Graf21_arXiv_eigenprojectors,Khanna97_AnnalsPhys_GeometricPF},
but in principle also facilitate development of new approaches to explore topological phenomena and quantum geometry at the forefront of physics research.

On another front, such a geometric characterization carries also the potential of catering to dynamically describing the evolution of quantum systems in larger Hilbert spaces~\cite{Hioe81_PRL}. For a two-level system, this is well established in terms of SO(3) rotations on the Bloch sphere. It is then a natural question to ask whether anything more can be harvested from the generalized rotations on the higher dimensional Bloch sphere for practical usage and what are the implications of these dynamics under the light of recent advances in controlling and manipulating quantum simulation experiments~\cite{Cooper19_RMP}.

We here answer these questions in the affirmative and show that such a geometric view of the SU($N$) dynamics not only unearths significant theoretical connections that has been previously overlooked but also bears important experimental consequences. We achieve this by employing generalized Gell-Mann matrices and a higher dimensional generalized Bloch hypersphere to describe $N$-level systems. While the quantum geometric tensor can be conveniently expressed on this hypersphere, in this paper we particularly focus on the paradigmatic Chern number in two spatial dimensions. Our main intuition lies at attributing a natural basis and a $\hat{z}$-axis role to the diagonal Gell-Mann matrices, which then allows us to construct a formalism to express the Chern number of an $N$-level system in terms of a collection of $N-1$ two-spheres. We demonstrate that the familiar Bloch sphere picture which depicts the Chern number in a two-level system indeed finds a correspondence for a multilevel system in the shape of an \textit{exterior sphere}. Within our framework, nontrivial topology manifests itself through band inversions on this exterior sphere governed by the last Gell-Mann matrix of the algebra (the basis is fixed by the formulation of the representation as described below). Moreover, this nested two-spheres picture immediately allows for new experimental probes such as the Wilson line measurement~\cite{Li16_Science_wilson} which can be easily extended to many-band systems to reveal the points where the state vectors are orthogonal to each other as a sign of nontrivial topology.

Secondly, by interpreting the time evolution of the spinor directly on the Bloch hypersphere we construct a state tomography protocol to extract the full coherence vector (the eigenvector) in a three-level system. Namely, we employ quenching techniques in cold atoms to project the state vector onto flat bands and investigate the dynamics of the coherence vector on the eight-dimensional hypersphere. We demonstrate that by analysing these generalized rotations which are restricted to certain planes dictated by the SU$(N)$ algebra, one can directly access the wave function.

%We promote the advantage of this geometric picture by constructing a tomography protocol for a three-level system which would allow for directly accessing the wave function and its dynamics on the generalized Bloch sphere in experiment.
%Quench protocols allow one to perform a state tomography for which we select \(N\)=3 and provide relations between coherence vector components.

The rest of the paper is organized as follows. We start in Sec.~\ref{sec: Gen. bloch sphere} by introducing our notation for the coherence vector by using the generalized Gell-Mann matrices and review the characterization of the Chern number on the Bloch sphere for two-level systems in Sec.~\ref{sec: two-levels} before expanding this geometric view to the multilevel case. In Sec.~\ref{sec: Nested_N=3} we prove that the Chern number of a three-level system can be expressed on two nested two-spheres and study the consequences of this description in detail by also employing example lattice models. Sec.~\ref{sec: N-levels} extends the nested spheres construction to the general $N$-level case where we also comment on how the exterior sphere can be probed via Wilson line measurements. We then turn to time evolution expressed in terms of generalized rotations directly on the Bloch hypersphere in Sec.~\ref{sec: tomography} and design a dynamical state tomography protocol for a three-level system by taking advantage of this framework. We conclude in Sec.~\ref{sec: Conclusion} with an outlook.

%%%%%%%%%%%%%%%%%%%%%%%%%%%%%%%%%%%%%%%%%%%%%%%%%%%%%%%%%%%%%%%%%%%%%%
\section{Generalized Bloch Hypersphere} \label{sec: Gen. bloch sphere}
%\subsection{The Coherence Vector}  \label{sec: Coh.vec.}
A generic $N$-level system can be described by using the $N^2-1$ (orthogonal) generators of the SU($N$) algebra (see Appendix~\ref{Appx:GellManns _andRotations}). These Hermitian and traceless operators [$\hat{\lambda}_j$ for integer $j=1,...,N^2-1$] can be chosen as the Pauli matrices ($\hat{\bm{\sigma}}$) for $N=2$ and the Gell-Mann matrices for $N=3$ levels. For larger Hilbert spaces generalized Gell-Mann matrices can be also systematically found for any $N$ (see e.g.~Ref.~\cite{Hioe81_PRL}). Any $N\times N$ Hermitian matrix can be expanded in the basis of the generalized Gell-Mann operators arranged as an ordered array, such as the Hamiltonian
\begin{equation} \label{eq:Hamiltonian vec}
\hat{H}=\bm{h}\cdot\bm{\hat{\lambda}} \,,
\end{equation}
or the density operator
\begin{equation}  \label{eq:rho def}
\hat{\rho}=\frac{1}{N}\hat{\id}_{N}+\frac{1}{2}\bm{S}\cdot\hat{\bs{\lambda}} \,,
\end{equation}
with vectors $\bm{h}$ and $\bm{S}$ defined in $\mathds{R}^{N^2-1}$. Here we choose a traceless Hamiltonian $\hat{H}$ without loss of generality by omitting the overall energy shift, whereas the density matrix $\hat{\rho}$ is ensured to have $\tr[\hat{\rho}]=1$. 

The density matrix associated to a quantum state can thus be equally expressed by using its corresponding \textit{coherence vector} $\bm{S}$, which is a generalized Bloch vector. Let us first elucidate some details of this notation. For a pure state ($\tr[\hat{\rho}^2]=1$), $\bm{S}$ lives on the surface of a hypersphere (i.e.~the generalized Bloch sphere) with a norm set by  $|\bm{S}|=\sqrt{2(N-1)/N}$, while mixed states feature a smaller length of  $\bm{S}$,   which can hence be seen as filling the interior of this hypersphere~\cite{Byrd07_IOP_CP2}. We mention that the coherence vector is sometimes defined in the literature with a different prefactor adjusted to have a unit norm also for a given $N>2$ with of course no effect on the dynamics. The definition in Eq.(\ref{eq:rho def}) is nonetheless generic for any $N$. In the following, we focus on pure states for the calculation of the Chern number, while noting that mixed states can be described in the same fashion for the time evolution dynamics. 

Eq.(\ref{eq:rho def}) essentially defines the coherence vector as the expectation value of the generalized Gell-Mann operators $\bm{S}=\tr[\hat{\rho}\hat{\bs{\lambda}}]=\langle\hat{\bs{\lambda}}\rangle$, where in principle $N^2-1$ such observables are needed. However, in reality there are further constraints on the wave function, hence, on the components of the coherence vector. This can be easily seen in the example of $N=3$ levels where the quantum state (qutrit) is associated to the complex projective plane $\bm{\mathds{C}\text{P}}^2=\text{SU}(3)/\text{U}(2)$~\cite{Byrd07_IOP_CP2}. In this case, a generic state is an element of $\mathds{C}^3$ as it requires three complex (six real) numbers, but the number of independent real parameters required to fully describe the pure state reduces to four upon imposing the normalization condition and choosing an overall phase (here we take the last component of the wave function to be real non-negative). Direct generalization to an $N$-level wave function that can be described by $N$ complex numbers ($2N$ real degrees of freedom) yields that one needs $2N-2$ real independent parameters to express a normalized state vector with fixed global phase.
%In general, an $N$-level wave function can be expressed by using $2N-2$ independent real parameters.

%Given the potency of the Bloch sphere picture in two levels, we here seek to generalize this geometric picture to larger Hilbert spaces. We are interested in the inner structure of the coherence vector $\bm{S}$ particularly relating to its topological properties and investigate the consequences of such geometric insight on the experiments. 

%%%%%%%%%%%%%%%%%%%%%%%%%%%%%%%%%%%%%%%%%%%%%%%%%%%%%%%%%%%%%%%%%%%%%%% 
%\subsection{Quantum Geometric Tensor}   \label{sec: QGT}
Motivation to more deeply investigate hidden connections involving this fundamental description of an $N$-level quantum state in terms of its coherence vector $\bm{S}(\bq)$ can be also obtained from the quantum geometric tensor (QGT)~\cite{Resta}. Concentrating on a state $\Psi_{\bq}$ separated by gaps from the rest of the spectrum in a given parameter space $\bq$ of dimension $D$, the QGT can be conveniently expressed in terms of the density matrix $\hat{\rho}(\bq)$ and the complementary projector $\hat{Q}(\bq)=\hat{\id}_{N}-\hat{\rho}(\bq)$ as $ \eta_{\alpha,\beta}(\bq)=g_{\alpha,\beta}(\bq)-\frac{i}{2}\Omega_{\alpha,\beta}(\bq)=\text{Tr}[\partial_{\alpha}\hat{\rho}(\bq)\hat{Q}(\bq)\partial_{\beta}\hat{\rho}(\bq)] $
for derivative $\partial_{\alpha}=\partial/\partial_{q_{\alpha}}$. The imaginary antisymmetric part $\Omega_{\alpha,\beta}(\bq)$ is nothing but the Berry curvature~\cite{Xiao10_RMP}, which, upon implementing the coherence vector definition in Eq.(\ref{eq:rho def}), can be associated with the anti-symmetric structure constant $f^{ijk}$ of the SU($N$) algebra via $\Omega_{\alpha,\beta}(\bq) =-\frac{1}{2} f^{ijk}S_{i}(\bq) \partial_{\alpha} S_j(\bq) \partial_{\beta} S_k(\bq)$~\cite{Graf21_arXiv_eigenprojectors,MooreVarma12_PRB_spin1}. Translating into the Chern number in $D=2$ dimensions, this allows for a geometric interpretation of the Chern number $C=\frac{1}{2\pi}\oint \dd^2q\, \Omega_{x,y}(\bq)$, directly on the generalized Bloch hypersphere
\begin{equation}   \label{eq:Chern_generalS}
        C=-\frac{1}{4\pi} \oint \dd^2q \, f^{ijk} S_{i}(\bq) \partial_{x} S_j(\bq) \partial_{y} S_k(\bq) \,,
\end{equation}
expanding the familiar expression (also given in Eq.(\ref{eq:Chern_2levelDefinion})) for the two-level case to the $N$-level systems. The real symmetric part $g_{\alpha,\beta}(\bq)$ corresponds to the quantum metric defining the `distance' between neighboring states and inherits the geometry of the underlying algebra through the two symmetric products featured in the SU($N$) group as detailed in Appendix \ref{Appx:GellManns _andRotations}.

These relations not only provide an alternative route for calculating the Berry curvature and the quantum metric, but also hint at the importance of a description in terms of the coherence vector that is deeply connected to the geometry of quantum states. Given the potency of the Bloch sphere picture in two levels, we here seek to generalize this intuitive geometric picture to larger Hilbert spaces by casting the diagonal SU($N$) generators as a natural basis. We are interested in the inner structure of the coherence vector $\bm{S}$ particularly relating to its topological properties such as the Chern number and investigate the consequences of such geometric insight on the experiments.

%%%%%%%%%%%%%%%%%%%%%%%%%%%%%%%%%%%%%%%%%%%%%%%%%%%%%%%%%%%%%%%%%%%%%%
\section{Two-Level System}    \label{sec: two-levels}
%The robust quantization of the transverse conductivity in the archetypal integer quantum Hall effect is known to be tied to the Chern number $C$. 
As  minimal models for Chern insulators, two-level systems admit a most useful geometric view in terms of the Bloch sphere. Therefore we first recapitulate this case to put the known characterization in the right perspective for developing our geometric description. 
We consider a periodic potential in two spatial dimensions with a gapped Hamiltonian spectrum consisting of two levels so that the energy eigenstates are well-defined everywhere. The cell periodic part of the Bloch states can be parameterized as $\ket{\Psi_{\bm q}}=(\cos(\theta_{\bm q}/2)e^{i\phi_{\bm q}},\sin(\theta_{\bm q}/2))^{\top}$ for crystal momentum $\bm q$. The corresponding coherence vector $\bm{S}({\bm q})=\bra{\Psi_{\bm q}}\hat{\bs{\lambda}}\ket{\Psi_{\bm q}}=(\sin\theta_{\bm q}\cos\phi_{\bm q}, \sin\theta_{\bm q}\sin\phi_{\bm q},\cos\theta_{\bm q} )^{\top}\equiv\bm{n}({\bm q})$ associates to unit two-sphere $\text{S}^2$, which we relabel as $\bm{n}$ to emphasize its nature of being the usual three-dimensional Bloch vector in spherical coordinates spanned by angles $(\theta_{\bm q},\phi_{\bm q})$. The Berry curvature of this state can be easily calculated via $\Omega_{\bm q}=-2\text{Im}\{\bra{\partial_{x}\Psi_{\bq}} \partial_{y}\Psi_{\bq} \ra \}$, with Im$\{.\}$ denoting the imaginary part, which gives $\Omega_{\bm q}=\frac{1}{2} J[\phi_{\bm q},\cos\theta_{\bm q}]$ for the Jacobian determinant $J$. Upon integrating across the Brillouin zone (BZ), one arrives at the Chern number which can also be expressed directly in terms of the Bloch vector as the 
Pontryagin formula~\cite{Pontryagin41},
\begin{equation}  \label{eq:Chern_2levelDefinion}
C=-\frac{1}{4\pi}\oint \dd^2q \, \bm{n} \cdot (\partial_{x}\bm{n} \times \partial_{y}\bm{n}) \,,
\end{equation} 
amounting for the solid angle covered by $\bm{n}(\bq)$ on the Bloch sphere (a special case of Eq.(\ref{eq:Chern_generalS}) with $f^{ijk}=\epsilon^{ijk}$). The Chern invariant notably equals the number of times the Bloch vector (or the Hamiltonian vector for that matter lying parallel to the eigenstates for $N=2$) wraps the Bloch sphere $\text{S}^2$, which is identified with $\pi_2(\text{S}^2)$ in homotopy language. In other words, when $C$ is nonzero, one cannot find a smooth unitary transformation that makes the Bloch vector point along the same direction e.g.~to the south pole everywhere, making the Berry curvature vanish at these points, in which case such a unitary transformation fails at the north pole where the entire contribution to the Chern number is left concentrated.

\begin{figure*}
	\centering\includegraphics[width=1\linewidth]{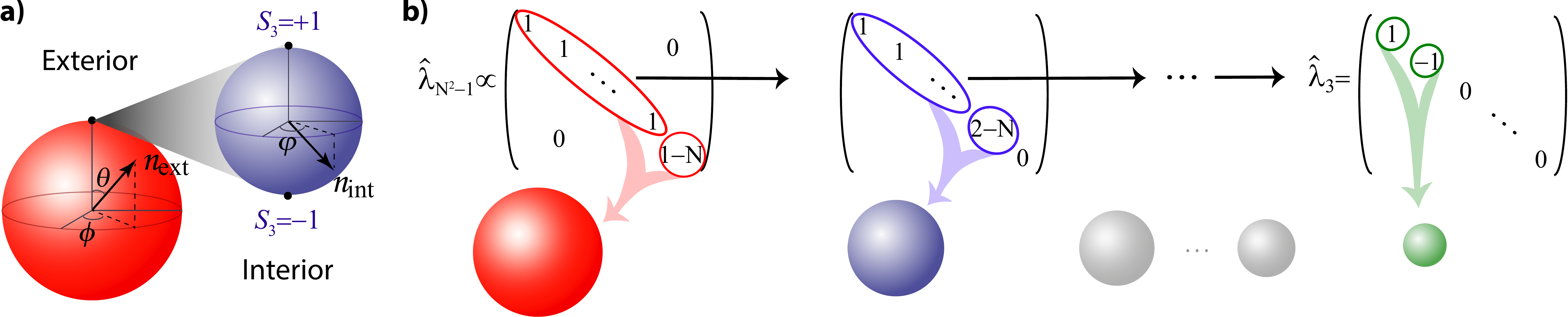}
	\caption{ Illustration of the nested two-spheres structure (a) in three-level systems. The exterior sphere (red) is a unit two-sphere spanned by the angles $(\theta,\phi)$ defined through Eq.(\ref{eq:Wf_3level_angles}), corresponds to mixing between the ground state and the rest of the bands via the last (diagonal) Gell-Mann matrix ($\hat{\lambda}_8$ in this case) upon expressing the wave function in an eigenstate basis as detailed in the text. The poles of the interior unit two-sphere (blue), associated with the mixing in the interior two-level via $\hat{\lambda}_3$, requires $\theta=0$ on the exterior one [Eq.(\ref{eq:S3-8definition_3level})], hence the name nested.
	(b) Generalization to $N$-levels harboring $N-1$ diagonal Gell-Mann matrices ($\hat{\lambda}_{k^2-1}$ for $2\leq k\leq N$). The $(k^2-1)^{\text{th}}$ Gell-Mann matrix consists of $k-1$ ones in its diagonal and one more nonzero term necessary to render it traceless. Similar to the diagonal Pauli matrix $\hat{\sigma}_z\rightarrow\hat{\lambda}_3$, these diagonal Gell-Mann matrices can be associated with the $\hat{z}$-axis of a two-sphere. When we start from the south pole of the exterior sphere, $\bm{n}_{\rm ext}$ vector must be inverted somewhere in the BZ and reach the north pole for a nontrivial Chern number.
	}
	\label{fig1_spheres}
\end{figure*}

However, this analogy is rather lost in multiband systems where the coherence vector is defined on a hypersphere that is hardly intuitive, and no such two-sphere characterization exists to date to the best of the authors' knowledge. Here we show that there is indeed a nested structure comprising $(N-1)$ two-spheres, among which an exterior sphere that plays a dominant role in controlling the Chern number of the state. We begin by expanding this geometric description first to three-level systems and then prove the nested spheres construction for any  number of bands, $N$. For this purpose, we consider the energy eigenstates of a lattice Hamiltonian and focus on the Chern number of the lowest band which is assumed to be separated from the rest of the spectrum with a gap so that the wave function and the Chern number are well-defined. We however note that the following discussion can be generalized to any state or band.

%%%%%%%%%%%%%%%%%%%%%%%%%%%%%%%%%%%%%%%%%%%%%%%%%%%%%%%%%%%%%%%%%%%%%
\section{Nested Two-Spheres for $N=3$ levels} \label{sec: Nested_N=3}
In this section, we imagine a two-dimensional lattice model consisting of three bands and consider the gapped ground state corresponding to a qutrit at each crystal momentum $\bm q$. We start by writing the wave function generically using four angles (rf.~Sec.\ref{sec: Gen. bloch sphere}) in a suggestive way,
\begin{equation} \label{eq:Wf_3level_angles}
    \ket{\Psi_{\bm q}}=\left(
    \begin{array}{ccc}
\cos(\theta_{\bm q}/2)\cos(\vartheta_{\bm q}/2)~e^{i(\phi_{\bm q}+\varphi_{\bm q})}\\
\cos(\theta_{\bm q}/2)\sin(\vartheta_{\bm q}/2)~e^{i(\phi_{\bm q}-\varphi_{\bm q})}\\
\sin(\theta_{\bm q}/2)
\end{array}\right) \,,
\end{equation}
where the angles \(0\leq[\theta,\vartheta]<\pi\) and \(0\leq[\phi,\varphi]<2\pi\) are respectively of polar and azimuthal nature. 
%The Berry connection and the Berry curvature of this state is defined via $\bm A_{\bm q}=i\braket{\Psi_{\bm q}|\bm{\nabla}\Psi_{\bm q}}$ and $\Omega_{\bm q}=\bm{\nabla}\times\bm{A}$ respectively. 
Laying the foundation of our two-sphere characterization, this form allows us to decouple the Berry curvature into two bilinear differential operators (i.e.~the Jacobian) as
\begin{equation}  \label{eq:BerryCurv_Jacobian_3level}
  \Omega_{\bm q}=\frac{1}{2}J\left[\phi_{\bm q},\frac{2}{\sqrt{3}}S_{8}(\bq)+\frac{1}{3}\right] +J\left[\varphi_{\bm q},S_{3}(\bq)\right] \,,
\end{equation}
in terms of the eighth $S_{8}(\bq)=\bra{\Psi_{\bm{q}}}\hat{\lambda}_8\ket{\Psi_{\bm q}}$ and the third components $S_{3}(\bq)=\bra{\Psi_{\bm q}}\hat{\lambda}_3\ket{\Psi_{\bm q}}$ of the coherence vector $\bm{S}(\bq)$,
\begin{eqnarray}  \label{eq:S3-8definition_3level}
-1\leq & \frac{2}{\sqrt{3}}S_{8}(\bq)+\frac{1}{3}=\cos\theta_{\bm q} &\leq1\,, \nonumber\\
  -1\leq& S_{3}(\bq)=\cos^2(\theta_{\bm q}/2)\cos\vartheta_{\bm q} &\leq1 \,.
\end{eqnarray}

Once articulated in this way, the $\hat{z}$-axis nature of the terms in Eq.(\ref{eq:S3-8definition_3level}) and their respective azimuthal angles depicted in Fig.\ref{fig1_spheres}a are stressed through the arguments of the Jacobians. Indeed, the first term in Eq.(\ref{eq:BerryCurv_Jacobian_3level}) is of the exact same form with the Berry curvature of a two-level system and represents the covering of a unit two-sphere spanned by $\{\phi_{\bm q},\cos\theta_{\bm q}\}$, which shall be henceforth dubbed the \textit{exterior sphere}. The second term is attributed to an \textit{interior sphere} $\{\varphi_{\bm q},S_3(\bm q)\}$, where such an overall polar angle definition would now involve a general function of both $\theta_{\bq}$ and $\vartheta_{\bq}$ through $S_{3}(\bq)$. Moreover, this interpretation allows us to define a three-dimensional vector $\bm{n}_{\rm ext}({\bm q})=(\sin\theta_{\bm q}\cos\phi_{\bm q}, \sin\theta_{\bm q}\sin\phi_{\bm q},\cos\theta_{\bm q} )^{\top}$ as in Fig.\ref{fig1_spheres}a, in terms of which the exterior sphere contribution can be expressed as $ J[\phi_{\bm q},\cos\theta_{\bm q}]=\bm{n}_{\rm ext} \cdot (\partial_{x}\bm{n}_{\rm ext} \times \partial_{y}\bm{n}_{\rm ext})$. Similarly, the interior sphere contribution can be related to the vector $\bm{n}_{\rm int}({\bm q})=(\sqrt{1-|S_3(\bm q)|^2}\cos\varphi_{\bm q}, \sqrt{1-|S_3(\bm q)|^2}\sin\varphi_{\bm q},S_3(\bm q) )^{\top}$.~We stress that these two nested two-spheres are highly correlated. In particular, while the north/south poles of the exterior sphere requires only $\theta_{\bm q}=0/\pi$ via Eq.(\ref{eq:S3-8definition_3level}), the interior sphere poles entail simultaneously $\theta_{\bm q}=0$ ($\bm{n}_{\rm ext}(\bq)=+\hat{z}$) and $\vartheta_{\bm q}=0/\pi$.

%In particular, Eq.(\ref{eq:BerryCurv_Jacobian_3level}) reveals the $\hat{z}$-axis characteristics of these components corresponding to the diagonal Gell-Mann matrices together with their associated azimuthal angles on a two-sphere. 
%Upon integration across the Brillouin zone (BZ) for the Chern number $C=\frac{1}{2\pi}\oint d^2q\,\Omega_{\bm q}$, the first term in Eq.(\ref{eq:BerryCurv_Jacobian_3level}) represents covering of the \([\phi,\frac{2}{\sqrt{3}}S_8+\frac{1}{3}]\) unit two-sphere (dubbed as the exterior sphere) and covering of the \([\varphi, S_3]\) unit two-sphere (an interior sphere). 
%the sum of which equals to the Chern number of the state as will be also confirmed numerically in the following?.

Note that the description of the wave function in Eq.(\ref{eq:Wf_3level_angles}) is not unique~\cite{Supp}. We, however, emphasize that our choice of representation is not coincidental, but in fact is rooted in the general form of the diagonal Gell-Mann matrices where each of them assumes a $\hat{z}$-axis role similar to the Pauli matrix $\hat{\sigma}_z$. We will demonstrate that $S_{3}(\bq)$ can be associated with the mixing on the interior two-level corresponding to $\hat{\lambda}_3=\text{diag}(1,-1,0)$, whereas $S_{8}(\bq)$ is a measure of the mixing between this nested interior two-level and the third component via $\hat{\lambda}_8=1/\sqrt{3}\,\text{diag}(1,1,-2)$. The latter effectively constitutes in its own right an exterior two-sphere on which the population difference is solely controlled by the angle $\theta_{\bm q}$. Elucidating the contributions to the Berry curvature in Eq.(\ref{eq:BerryCurv_Jacobian_3level}) as coming from these exterior and interior two-spheres will have important implications for the Chern number of the state, topological singularities and band inversion(s) necessary to have non-trivial topology.

\begin{figure}
	\centering\includegraphics[width=1\linewidth]{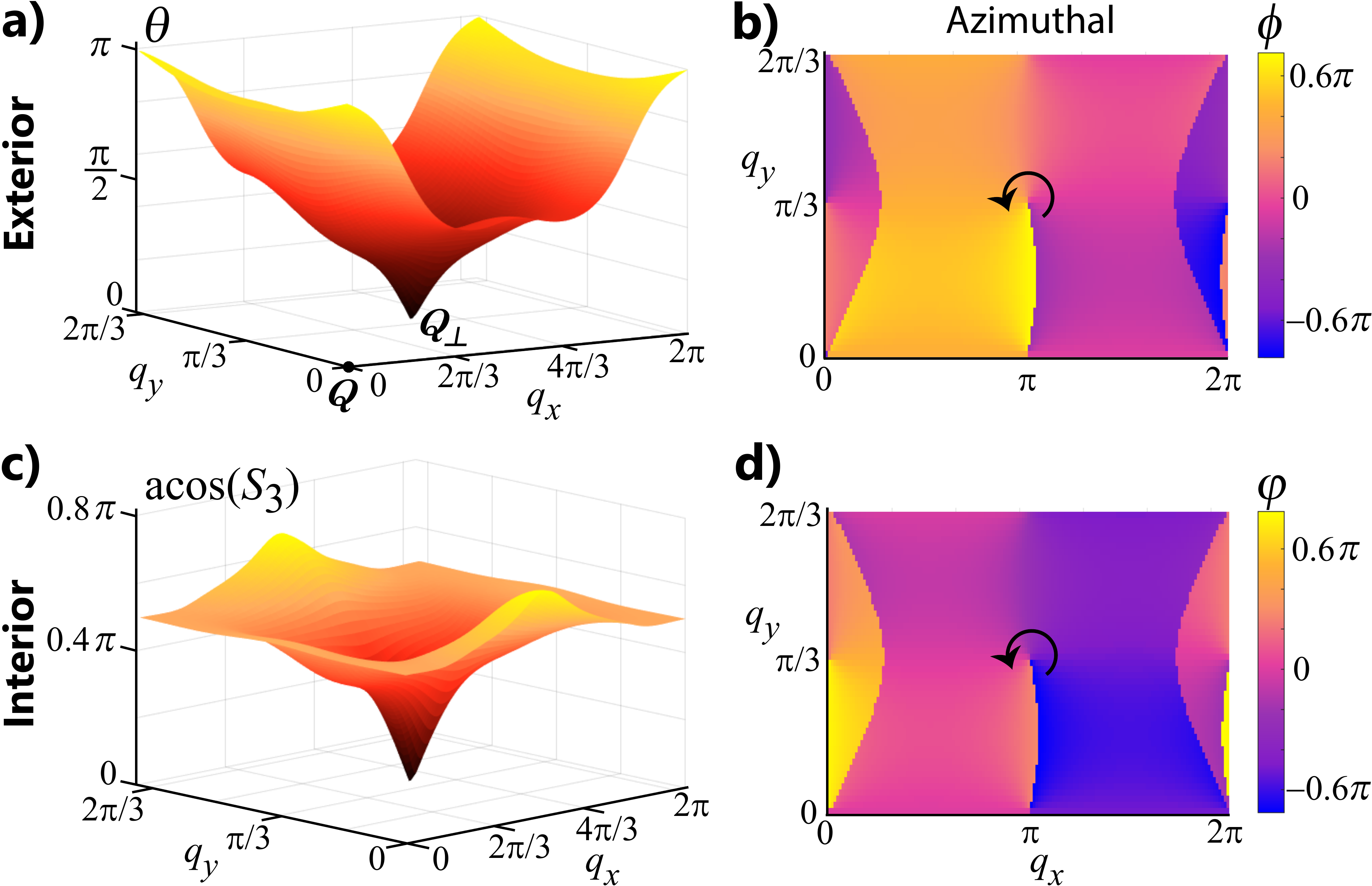}
	\caption{ Distributions on the exterior (top panel) and interior (bottom) two-spheres for the ground state of the Harper-Hofstadter model (\ref{eq:HofstHamiltonian}) with reference to the eigenstates at ${\bm Q}=\Gamma$, as a function of crystal momentum $\bq$ with lattice constant set to one. (a) Starting from the exterior south pole ($\theta_{\bm Q}=\pi$), the state vector reaches the exterior north pole ($\theta_{\bm{Q}_{\perp}}=0$) once within the BZ, a prerequisite for nontrivial $C$, (c) while simultaneously the interior vector is found to be on its north pole $S_3(\bm{Q}_{\perp})=+1$. The total winding of the respective azimuthal angles $\Delta(\phi+\varphi)=2\pi$ at this band inversion point gives $C=1$.
	}
	\label{fig2_ExtInt.Angles}
\end{figure}

Specifically, we consider the eigenstates at some reference point $\bm{Q}$ in the BZ as our basis choice, which we label as $\{\ket{1},\ket{2},\ket{3}\}$ ordered from the lowest to the highest in energy. We write the ground state wave function at any other momenta ${\bm q}$ in this eigenstate basis as
$ \ket{\Psi_{\bm q}}=\cos(\theta_{\bm q}/2)\cos(\vartheta_{\bm q}/2)~e^{i(\phi_{\bm q}+\varphi_{\bm q})} \ket{3} + \cos(\theta_{\bm q}/2)\sin(\vartheta_{\bm q}/2)~e^{i(\phi_{\bm q}-\varphi_{\bm q})} \ket{2} + \sin(\theta_{\bm q}/2) \ket{1}$.
%in such a way that the rows of the wave function given in Eq.(\ref{eq:Wf_3level_angles}) correspond to the overlap with the basis vectors in decreasing order (i.e.~$\la{1}\ket{\Psi_{\bq}}=\sin(\theta_{\bq}/2)$). 
We now analyze the distribution of $\bm{n}_{\rm ext}(\bq)$ and $\bm{n}_{\rm int}(\bq)$ vectors across the BZ on their respective spheres. At the reference point $\ket{\Psi_{\bm{Q}}}=\ket{1}=(0,0,1)^{\top}$ with $\theta_{\bm{Q}}=\pi$, we therefore start from the south pole on the external sphere $\bm{n}_{\rm ext}(\bm{Q})=-\hat{z}$, while $\bm{n}_{\rm int}(\bm{Q})$ lies in the equator with $S_3(\bm{Q})=0$. Moving away from the reference point, $\bm{n}_{\rm ext}(\bq)$ vector begins to deviate from the south pole but can be rotated back to it with a smooth unitary transformation along the shortest great circle connecting these points similar to the case of a two-level system. This procedure fails only if $\bm{n}_{\rm ext}(\bm{Q}_{\perp1})=+\hat{z}$ is at the north pole ($\theta_{\bm{Q}_{\perp1}}=0$) for some $\bm{Q}_{\perp1}$, for which the required rotation matrix is not uniquely defined on the external sphere. 

A topologically nontrivial state with $C\neq0$ can be defined as a state that cannot be smoothly connected to the atomic limit, for which instead band inversions are required. $\theta_{\bm{Q}_{\perp1}}=0$ accomplishes exactly this by inverting the band character with respect to the reference state on the external sphere and the wave function becomes orthonormal to the reference state $\bra{\Psi_{\bm{Q}}} \Psi_{\bm{Q}_{\perp1}} \ra=0$. 
Since the Berry curvature can be made to vanish smoothly with a unitary transformation everywhere else, we infer that, in order to have a nontrivial Chern number, $\bm{n}_{\rm ext}(\bq)$ has to visit both poles of the exterior sphere.
We now analyze possible cases that can occur:
%both $\ket{\Psi_{\bm{Q}}}=(0,0,1)^{\top}$ and $\ket{\Psi_{\bm{Q}_{\perp1}}}=(a,b,0)^{\top}$ within the BZ for some complex $a,b$.

\textbf{Case 0:} Without $\theta_{\bq}$ vanishing, the internal sphere does not have the opportunity to be fully covered and can be nullified like the external one. Therefore, if such $\bm{Q}_{\perp}$ is not present, $\Psi_{\bq}$ can be smoothly deformed to $(0,0,1)^{\top}$ everywhere, corresponding to a trivial Chern number with vanishing Berry curvature following the smooth deformation (see also Sec.~\ref{Sec:U_q}).

\begin{figure}
	\centering\includegraphics[width=1\linewidth]{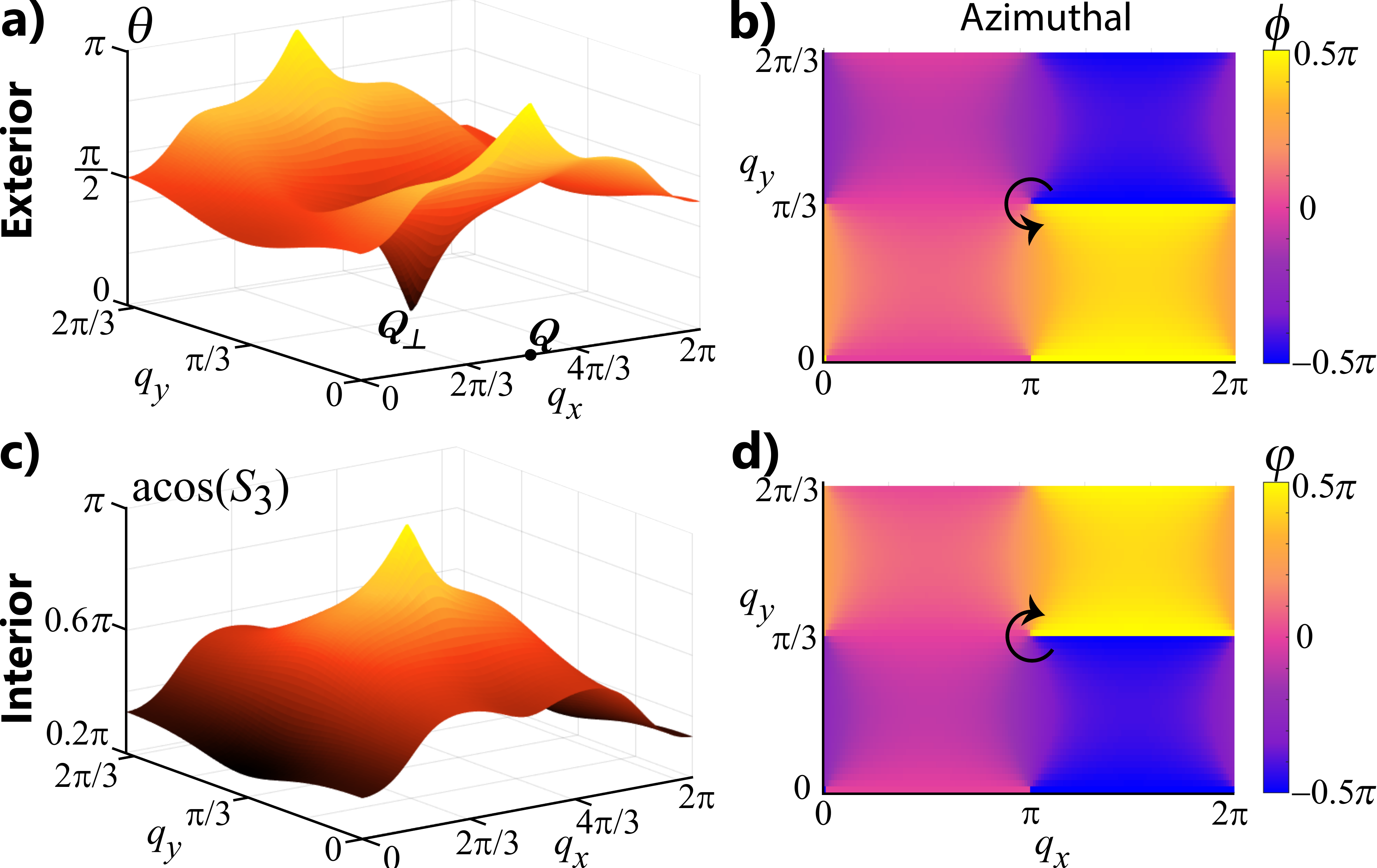}
	\caption{ Similar to Fig.~\ref{fig2_ExtInt.Angles} but for ${\bm Q}=(\pi,0)$. (a) The vector $\bm{n}_{\rm ext}(\bq)$ gets inverted on the exterior sphere again once, visiting both poles. (c) $S_3(\bq)$ can reach one of the interior sphere poles at this point, here $S_3(\bm{Q}_{\perp})=-1$ is at the south pole. (b,d) The exterior sphere azimuthal angle $\phi$ winds by $\pi$ around this singularity (black arrow), while the interior winding is $-\pi$ but around the south pole, adding up to $C=1$ via Eq.(\ref{eq:WindingAroundSing_ChernNum}).
	}
	\label{fig3_Q=(pi,0)_ExtInt.Angles}
\end{figure}

\textbf{Case 1:} If there is only one band inversion in the BZ with $\theta_{\bm{Q}_{\perp1}}=0$ (as in Figs.~\ref{fig2_ExtInt.Angles} and~\ref{fig3_Q=(pi,0)_ExtInt.Angles}), we can calculate the winding of the U(1) phase singularity around this point by using the Berry connection, ${\bm A}_{\bq}=-\cos^2(\theta_{\bq}/2)\bm{\nabla}\phi_{\bq} - \cos^2(\theta_{\bq}/2)\cos(\vartheta_{\bq})\bm{\nabla}\varphi_{\bq}$, of the state given in Eq.(\ref{eq:Wf_3level_angles}). We imagine a closed area $\mathcal{B}$ enclosing the singularity and its boundary $\partial\mathcal{B}$. In the limit of $\mathcal{B}\rightarrow0$, the Chern number $C$ can be related to the line integral of the Berry connection around the singularity,
\begin{equation}
\oint\limits_{\partial\mathcal{B}} \dd\bq \cdot{\bm A}_{\bq} = -\oint\limits_{BZ-\{\mathcal{B}\}} \dd^2q \,(\bm{\nabla}\times{\bm A}_{\bq})\cdot\hat{z}=-2\pi C \,,
\end{equation}
where the surface integral is taken over the entire BZ excluding the singularity.
Close to this singularity point, the wave function takes the form $ \ket{\Psi_{\bq}}=(\cos(\vartheta_{\bq}/2)~e^{i(\phi_{\bq}+\varphi_{\bq})},\sin(\vartheta_{\bq}/2)~e^{i(\phi_{\bq}-\varphi_{\bq})},0)^{\top}$. Therefore, singularities can appear in $\phi$ if $\vartheta\neq0,\pi$; in  $\phi+\varphi$ if $\vartheta=0$; and in $\phi-\varphi$ if $\vartheta=\pi$. This translates into the Chern number via the Berry connection formula as
\begin{equation} \label{eq:WindingAroundSing_ChernNum}
    C=w=\frac{1}{2\pi} \Bigg\{
    \begin{array}{ccc}
&\Delta\phi,            &\theta=0,\vartheta\neq0,\pi \,,\\
&\Delta(\phi+\varphi),  &\theta=0,\vartheta=0 \,,\\
&\Delta(\phi-\varphi),  &\theta=0,\vartheta=\pi \,,
\end{array}
\end{equation}
where $\Delta\phi$ is the winding of the phase $\phi$ around the singularity $\partial\mathcal{B}$ and similarly for the others. The interior sphere pole visited by the state (i.e.~$\sgn(\cos\vartheta)$) determines the sign of the winding of its azimuthal angle $\varphi$. Note that this discussion can be easily extended to the Chern number of higher bands since the results for the Berry connection and curvature are generic\footnote{The wave function of the $n^{\text{th}}$ (isolated) band is expressed in the eigenstate basis at some reference momentum $Q$ in such way that its last element corresponds to the distribution within that band, i.e.~$\la\Psi^n_q|\Psi^n_Q\ra=\sin\theta_q$. Similarly, $|\Psi^n_q\ra$ must become orthogonal with respect to the reference point, corresponding to the poles of the exterior sphere.}.
%This essentially means that if the internal vector is also at one of its poles $|\bm{n}_{\rm int}|=\hat{z}$, the contribution of the winding of $\varphi$ happens in a way that does not annihilate the winding of the $\phi$-angle on the external sphere but instead complements it. This is easy to see upon imagining a situation where the ground state wave function is inverted \textit{once} within the BZ, hence mixed with the other bands, but still corresponds to a trivial Chern band which is contradictory to the band inversion. 

\textbf{Case 2:} Should there exist multiple singularity points ($j=1,..,N_s$) with $\theta_{\bm{Q}_{\perp j}}=0$, winding contributions coming from each singularity must be added for the Chern number $C=\sum_j w_j$, where the local winding numbers $w_j$ are defined as in Eq.(\ref{eq:WindingAroundSing_ChernNum}). Note that it is possible for local contributions to cancel each other. However, in physical examples we looked at there is usually only a single point ${\bm{Q}_{\perp}}$ for $C=1$, while we find multiple singularities when some special high-symmetry points are chosen as reference or for higher Chern numbers that naturally require more twisting.
%However, multiple singularities in general require more twisting in the wave function or special points in the BZ chosen for the reference states, and we observe that for $C=1$ there is usually only a single point ${\bm{Q}_{\perp}}$ contributing to the Chern number. 
We now illustrate these cases in more detail by implementing specific lattice models.

\subsection{Application to Lattice Models}
We consider the lowest band ($C=1$) of the Harper-Hofstadter model at $1/3$ flux-quantum per plaquette in a square lattice, with the Hamiltonian
\begin{equation}  \label{eq:HofstHamiltonian}
        \hat{H}=-t\sum_{j,m}(e^{2\pi mi/3}\hat{a}_{j+1,m}^{\dagger}\hat{a}_{j,m}+\hat{a}_{j,m+1}^{\dagger}\hat{a}_{j,m}+h.c.) \,,
\end{equation}
written in the Landau gauge, where  $\hat{a}_{j,m}^{(\dagger)}$ annihilates (creates) a particle at lattice site labeled with $(j,m)$ with lattice constant ($a=1$) set to one, $t$ is the tunneling strength and $h.c.$ stands for the hermitian conjugate~\cite{Cooper19_RMP}. We numerically find the eigenstates at some reference momentum $\bm{Q}$ and analyze the distributions on the external and internal spheres across the BZ by expressing the ground state wave function $\Psi_{\bq}$ in the basis of the reference eigenstates. We plot the polar angles defined through Eq.(\ref{eq:Wf_3level_angles}) on the two-spheres together with their azimuthal counterparts in Fig.\ref{fig2_ExtInt.Angles} and Fig.\ref{fig3_Q=(pi,0)_ExtInt.Angles} for $\bm{Q}=\bm{\Gamma}$ and $\bm{Q}=(\pi,0)$ respectively \footnote{Note that for the reference state choice the only requirement is that the ground state should be gapped, as it should be everywhere else across the BZ to be able to define a proper band inversion and Chern number. Excited states at the reference momentum can be chosen to be degenerate, which we find can help with smoothing the azimuthal angles in numerics without changing the $\theta$ distribution. The azimuthal angle distributions can be also smoothed by allowing $\pm\pi$ addition to both, under which the wave function (\ref{eq:Wf_3level_angles}) remains unchanged.}. It can be clearly seen that the ground state wave function becomes orthogonal to itself once in the middle of the BZ (${\bm{Q}_{\perp}}$) in both figures, hence visiting both poles of the external sphere.

While for most reference points the internal sphere poles are not reached and contribution to the Chern number comes entirely from the winding in $\phi$, we choose these parameters to display also the role of the internal sphere poles that is complementary to the external one. In Fig.\ref{fig2_ExtInt.Angles}c, ${\bm n}_{\rm int}({\bm{Q}_{\perp}})=+\hat{z}$ reaches its north pole, corresponding to the second line in Eq.(\ref{eq:WindingAroundSing_ChernNum}). With the winding in $\phi$ and $\varphi$ both counterclockwise (black arrows in Fig.\ref{fig2_ExtInt.Angles}(b,d)), they add up to $\Delta(\phi+\varphi)=2\pi$ hence $C=1$. On the other hand in Fig.\ref{fig3_Q=(pi,0)_ExtInt.Angles}, ${\bm n}_{\rm ext}({\bm{Q}_{\perp}})=+\hat{z}$ is at the north pole, while ${\bm n}_{\rm int}({\bm{Q}_{\perp}})=-\hat{z}$ is at the south pole (i.e.~$\theta=0,\vartheta=\pi$) where the winding in the internal azimuthal angle $\varphi$ is also reversed to clockwise. This however is not a problem but exactly what is required, since the Chern number
is given by the difference of the two azimuthal angles in Eq.(\ref{eq:WindingAroundSing_ChernNum}) where $\Delta(\phi-\varphi)=2\pi$ and $C=1$. We numerically confirm this is indeed true generically and the ground state vector gets inverted on the external sphere for different $\bm{Q}$ values chosen along the high symmetry lines or in random points in the BZ in the Harper-Hofstadter model, as well as for the ground state of the Lieb lattice with $C=1$~\cite{PalumboMeichanetzidis15_PRB}, and the triangular optical flux lattices with $C=1$ and $C=2$~\cite{CooperMoessner12_PRL,CooperDalibard13_PRL}, signaling nontrivial topology in all of the models.

%This essentially means that if the internal vector is also at one of its poles $|\bm{n}_{\rm int}|=\hat{z}$, the contribution of the winding of $\varphi$ happens in a way that does not annihilate the winding of the $\phi$-angle on the external sphere but instead complements it. This is easy to see upon imagining a situation where the ground state wave function is inverted \textit{once} within the BZ, hence mixed with the other bands, but still corresponds to a trivial Chern band which is contradictory to the band inversion. 

\begin{figure}
	\centering\includegraphics[width=1\linewidth]{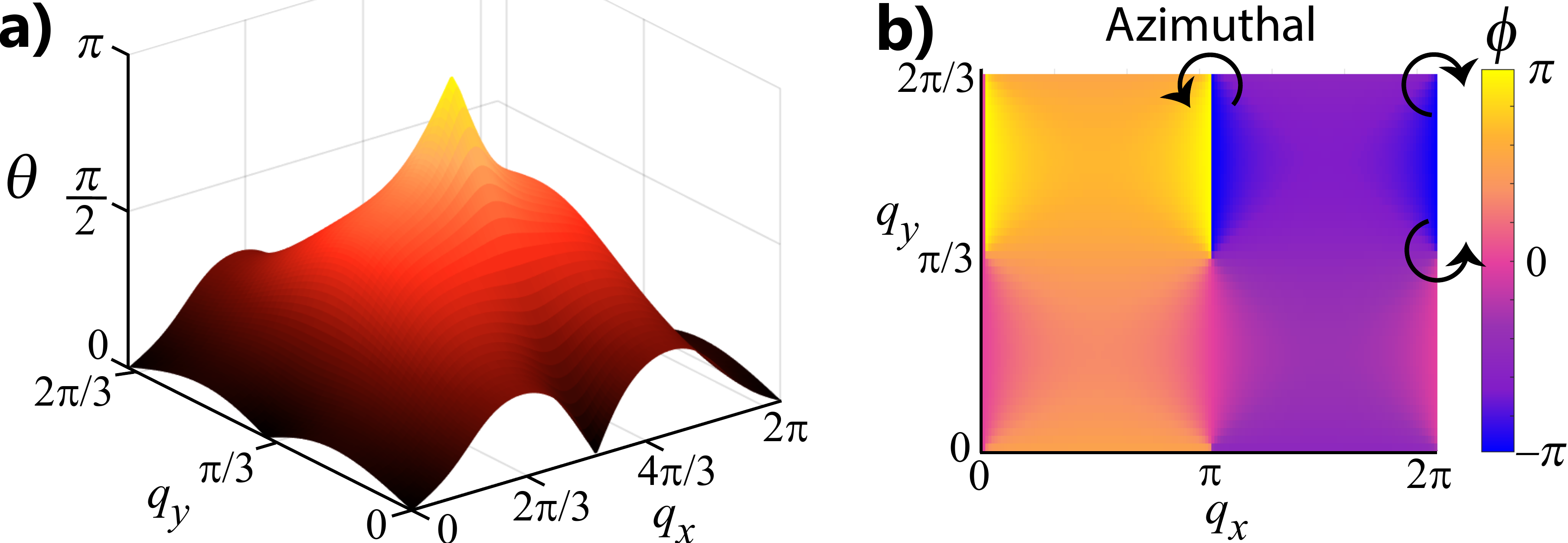}
	\caption{ Polar and azimuthal angles on the exterior sphere in the Harper-Hofstadter model (\ref{eq:HofstHamiltonian}). (a) The ground state wave function becomes orthogonal three times with respect to the eigenstate at ${\bm Q}=(\pi,\pi/3)$; at $\Gamma$ and two M points where $\theta(\bq)$ vanishes. (b) Adding the winding of $|\Delta\phi|=2\pi$ around these singularities, one of which has opposite chirality, we find $C=1$. Here, also the entire contribution to the Chern number comes from the exterior sphere, where the interior sphere azimuthal angle is found to be constant and $|S_3(\bq)|<1$ everywhere.}
	\label{fig4_Ext_multipleBandInv}
\end{figure}

Although, for $C=1$, this band inversion defined on the external sphere usually occurs only once within the BZ, we observe that there are multiple singularities (Case 2) with respect to the ground state wave function at $\bm{Q}=(\pi,\pi/3)$ in the Harper-Hofstadter model~(\ref{eq:HofstHamiltonian}). The external sphere angles are demonstrated in Fig.\ref{fig4_Ext_multipleBandInv} where we find that $\varphi_{\bq}=0$ is fixed and the internal sphere poles are not reached [54]. There are three singularities where $\theta$ vanishes at ${\bm{Q}_{\perp1}}=\Gamma$, ${\bm{Q}_{\perp2}}=(0,\pi/3)$ and ${\bm{Q}_{\perp3}}=(\pi,0)$, each resulting in a phase winding of $|\Delta\phi|=2\pi$. The chirality of them though are $-1,+1,+1$; adding up to the correct value $C=1$. We note that the chirality at the singularity points can be also found by defining the sign of the Jacobian $Ch=\sum_{\bm{Q}_{\perp j}} \text{sgn}[\hat{z} \cdot (\partial_{x}\bm{n}_{\rm ext} \times \partial_{y}\bm{n}_{\rm ext})]$ similar to the Brouwer degree of the map to the Bloch sphere in the two-level case~\cite{SticletSimon12_PRB_brouwer}. Care must be given while calculating this Jacobian which must be nonzero. If the Jacobian is instead associated to a quadratic band touching point and vanishes, another reference point should be chosen~\cite{SticletSimon12_PRB_brouwer}.

Within our framework, a nontrivial contribution to the Chern number can indeed originate only from the poles of the two-spheres corresponding to the Case 1 and Case 2 given in the previous subsection,
%beginning of Sec.~\ref{sec: Nested_N=3},
for which $\theta_{\bm{Q}_{\perp}}=0$ is compulsory~\footnote{In our parametrization, the Berry connection is zero at the starting point i.e.~the exterior south pole and Eq.(\ref{eq:Wf_3level_angles}) corresponds to the wave function for the southern hemisphere with the singularity at the north pole.}. Upon taking the line integral of the Berry connection around these band inversion points, winding of the external azimuthal angle $\phi_{\bq}$ (complemented by the internal azimuthal angle if $\bm{n}_{\rm int}(\bm{Q}_{\perp})=\pm\hat{z}$ is also at one of its poles) successfully reproduces the Chern number. Nonetheless, one must also analyze consequences of $\vartheta_{\bq}$ becoming zero (or $\pi$) without vanishing $\theta_{\bq}$. In fact, these are exactly the points where we observe additional windings in the individual azimuthal angles in Figs.~\ref{fig2_ExtInt.Angles} and \ref{fig3_Q=(pi,0)_ExtInt.Angles} e.g.~at $\bq=(0,\pi/3)$ outside of the poles, whose total contribution though vanishes as the continuity of the coherence vector is ensured. For example, winding of the Berry connection around an infinitesimal area surrounding any $\bq_i$ with $(\theta_{\bq_i}\neq0,\vartheta_{\bq_i}=0)$ can be written as $\oint_{\partial\mathcal{B}} \dd\bq \cdot{\bm A}_{\bq}=-\oint_{\mathcal{B}} \dd^2q \cos^2(\theta_{\bq}/2) \bm{\nabla}\times\bm{\nabla}(\phi_{\bq}+\varphi_{\bq})$, which results in a nonzero contribution of $-\cos^2(\theta_{\bq}/2)\delta(\bq-\bq_i)$ only if $(\phi_{\bq_i}+\varphi_{\bq_i})$ is singular. One can however see that this is not allowed upon considering that the coherence vector must be continuous for a smooth Hamiltonian, particularly the fourth and fifth components given as $[S_4(\bq_i),S_5(\bq_i)]=[\sin\theta_{\bq_i}\cos(\phi_{\bq_i}+\varphi_{\bq_i}), \sin\theta_{\bq_i}\sin(\phi_{\bq_i}+\varphi_{\bq_i}) ]$ for these parameters. Similarly, the sixth and seventh components of $\bm{S}(\bq)$ carry the term $(\phi_{\bq_i}-\varphi_{\bq_i})$, inhibiting a singularity in the difference of these angles for the case of $(\theta_{\bq_i}\neq0,\vartheta_{\bq_i}=\pi)$. In physical examples we study, we find that there can be windings in individual $\phi_{\bq}$ and $\varphi_{\bq}$ profiles outside of the poles as is visible in figures, which we attribute to external and internal spheres being highly intertwined, but we numerically confirm that their winding contributions cancel each other as expected.

%Notably, in the Hofstadter model, if the middle of the shown BZ with $\bm{Q}=(\pi,\pi/3)$ is chosen as the reference, we find that the wave function gets inverted on the external sphere at three different places (at $\bm{\Gamma}$ and two $\bm{M}$ points) as demonstrated in Fig.\ref{fig4_Ext_multipleBandInv}. Chirality at these points is nonetheless found to be $+1,-1,+1$; adding up to the correct value of $C=1$. 

\subsection{Local Unitary Transformation} \label{Sec:U_q}
The total contribution of the external and internal spheres can be found by integrating the terms in Eq.(\ref{eq:BerryCurv_Jacobian_3level}) individually, while only their sum is fixed and the quantized Chern number. In principle, one can find a reference point $\bm{Q}$ where the entire Berry curvature emanates from the external sphere with a vanishing contribution from the internal structure. Indeed, this is exactly the case for the parameters given in Fig.\ref{fig4_Ext_multipleBandInv}, where we numerically calculate $C=\frac{1}{4\pi}\oint \dd^2q\, \bm{n}_{\rm ext} \cdot (\partial_{x}\bm{n}_{\rm ext} \times \partial_{y}\bm{n}_{\rm ext})=0.998$. However, this is typically not valid and when a different eigenstate basis ($\bm{Q}$) is chosen, external and internal sphere contributions change. It is nonetheless possible to nullify the contribution of the inner sphere by applying a local unitary transformation~\footnote{Note that locally the Berry curvature can in principle change upon unitary transformation of this form as with the Haldane model, but the total Chern number remains the same~\cite{Cooper19_RMP}. See also~\cite{BenaMontambaux09_NJP}} of the form $\hat{U}_{\bq}=\text{exp}\{-if(\bq)\hat{\sigma}_z\}\bigoplus\text{exp}\{0\}$ to the state~(\ref{eq:Wf_3level_angles}), which unwinds the internal azimuthal angle $\varphi$ when $f(\bq)=\varphi_{\bq}$ by acting on the internal structure. 

Under the action of this local unitary transformation, the total Berry curvature (\ref{eq:BerryCurv_Jacobian_3level}) is modified $\Omega'_{\bq}=\frac{1}{2}J[\phi_{\bm q},\cos\theta_{\bm q}]+S_3(\bq)\bm{\hat{z}}\cdot(\bm{\nabla}\times\bm{\nabla}\varphi_{\bq})$ to account for the singularities in $\varphi_{\bq}$, while the contribution of the external sphere remains unchanged. Note that if $\varphi_{\bq}$ is smoothly varying as a function of $\bq$, curl of the gradient here is automatically zero, in which case entire contribution to the Chern number comes from the external sphere as in Fig.\ref{fig4_Ext_multipleBandInv}. However, if singularities are present at $\varphi_{\bq_j}$, upon integration the second term in $\Omega'_{\bq}$ contributes as $C=\frac{1}{4\pi}\oint \dd^2q\,J[\phi_{\bm q},\cos\theta_{\bm q}] +\frac{1}{2\pi} \sum_j S_3(\bq_j) \Delta\varphi_{\bq j}$, in agreement with Eq.(\ref{eq:WindingAroundSing_ChernNum}). We numerically confirm that this vortex term carries the precise amount necessary to top up the external sphere contribution to $C=1$ in our simulations for the Harper-Hofstadter model.

Furthermore, this local unitary transformation also caters to illustrating the `Case 0' which corresponds to a trivial invariant with no band inversions on the external sphere. Indeed, if there are no points present in the BZ where $\theta_{\bq}$ vanishes, the state vector does not reach the north pole on the external sphere or either of the poles on the internal one. Upon applying $\hat{U}_{\bq}$, no topological winding contribution comes from the internal sphere and the Chern number reduces to the external sphere contribution. The external sphere integral though vanishes as well since the ${\bm n}_{\rm ext}(\bq)$ vector can be rotated to point to the south pole everywhere smoothly with another unitary transformation as it never reaches the north pole.

%\note{might need to discuss more?: spheres are highly intertwined, the azimuthal angles, their ranges...}

%%%%%%%%%%%%%%%%%%%%%%%%%%%%%%%%%%%%%%%%%%%%%%%%%%%%%%%%%%%%%%%%%%%%%
\section{General $N$ levels}  \label{sec: N-levels}
Our nested spheres construction readily extends to \(N\)-level Chern insulating systems, where there are (\(N-2\)) nested interior structures embedded in the exterior sphere as in a set of \((N-1)\) Russian dolls of two-spheres whose $\hat{z}$-axes are associated to the diagonal Gell-Mann matrices as illustrated in Fig.\ref{fig1_spheres}b. We prove this by starting from the ground state wave function and write it generically in terms of some amplitudes and phases, 
\begin{equation}   \label{eq:Nlevel-WaveFunc-amplitudes}
\ket{\Psi_{\bm q}}=\left(
    \begin{array}{cccc}
    \lvert a_1(\bq)\rvert e^{i\alpha_1(\bq)}  \\
     \lvert a_2(\bq)\rvert e^{i\alpha_2(\bq)}\\
     \vdots\\
     \lvert a_N(\bq)\rvert
\end{array}\right),
\end{equation}
where we choose $\alpha_N=0$. The Berry curvature, $\Omega_{\bm q}=\bm{\hat{z}}\cdot(\bm{\nabla}\times\bm{A}_{\bm q})$, of this state can be easily found by taking the curl of the Berry connection $\bm A_{\bm q}=i\braket{\Psi_{\bq}|\bm{\nabla}\Psi_{\bq}}=-\sum_{j=1}^{N-1}\lvert a_j(\bq)\rvert^2\bm{\nabla}\alpha_j(\bq)$, which gives
\begin{equation}  \label{eq: Nlevel-Jacobians-amplitudes}
\Omega_{\bm q}=\sum_{j=1}^{N-1}J[\alpha_j,\lvert a_j\rvert^2]=\sum_{j=1}^{N-1}\frac{1}{2}J[\alpha_j,2\lvert a_j\rvert^2-1] \,.
\end{equation}
Upon rearranging the coefficients, the Jacobians can now be associated to two-spheres whose $\hat{z}$-axes are in the correct range $-1\leq2\lvert a_k\rvert^2-1<1$, with azimuthal angles $0\leq\alpha_k<2\pi$. Although this form is general and can be also implemented in the calculation of the total Berry curvature, we further readjust these terms to formulate them in terms of the coherence vector components, which then directly describe population inversion within each nested structure (Fig.\ref{fig1_spheres}b). 

SU$(N)$ algebra harbors $N-1$ diagonal Gell-Mann matrices whose associated coherence vector components for the wave function (\ref{eq:Nlevel-WaveFunc-amplitudes}) take the form
\begin{equation}
    S_{k^2-1}(\bq)=\sqrt{\frac{2}{k(k-1)}}\left(\sum_{j=1}^{k-1}\lvert a_j(\bq)\rvert^2-(k-1)\lvert a_k(\bq)\rvert^2\right)
\end{equation}
for \(2\leq k\leq N\). Note though that only \(S_3(\bq)\) is canonically normalised, all other components must be re-normalised as,
\begin{equation}
    -1\leq\sqrt{\frac{2(k-1)}{k}}S_{k^2-1}(\bq)+\frac{k-2}{k}\leq1 \,,
\end{equation}
which generalizes Eq.(\ref{eq:S3-8definition_3level}) to any dimensions.
Our goal is to collect the amplitude terms in Eq.(\ref{eq: Nlevel-Jacobians-amplitudes}) in such a way that the diagonal coherence vector components are rendered into the second argument of the Jacobians. This can be achieved by redefining the phases for \(1\leq k\leq N-1\) as
\begin{equation}
    \alpha_k(\bq)=\sum_{j=k}^{N-1}\phi_j(\bq)-(k-1)\phi_{k-1}(\bq) \,,
\end{equation}
where the new azimuthal angles $\phi_k(\bq)$ remain decoupled. The Berry curvature can be then written as
\begin{eqnarray}  \label{eq:BerryCurv_Jacobians_anyN}
\Omega_{\bm q}&=&\frac{1}{2}J\left[\phi_{N-1},\sqrt{\frac{2(N-1)}{N}}S_{N^2-1}+\frac{N-2}{N}\right] \nonumber \\
\!\!&&\!\!\!\! +\!\! \sum_{k=2}^{N-1}\frac{k}{2}J\left[\phi_{k-1},\sqrt{\frac{2(k-1)}{k}}S_{k^2-1}+\frac{k-2}{k}\right]\!\!.
\end{eqnarray}

This form expands the Berry curvature that is commonly expressed as the solid angle on a two-sphere for two-level systems and that we established in terms of two nested two-spheres for the three-level case, to systems with any number of bands $N\geq2$ as illustrated in Fig.\ref{fig1_spheres}b. Indeed, the first Jacobian in Eq.(\ref{eq:BerryCurv_Jacobians_anyN}) again corresponds to the wrapping of an exterior sphere associated with the $(N^2-1)^{\text{th}}$ Gell-Mann matrix. Upon considering an eigenstate basis, this last Gell-Mann matrix, $\hat{\lambda}_{N^2-1}\propto\text{diag}[1,1,\ldots1,-(N-1)]$, underpins the mixing of the the ground state with the collection of the excited states. Similarly, the $N$-level state vector (\ref{eq:Nlevel-WaveFunc-amplitudes}) can be expressed in terms of $N-1$ azimuthal angles $\phi_k$ and polar angles defined in a nested way. We present the general form of the four and five-level wave functions in Appendix~\cite{Supp}. In particular, the last element of the wave function ($a_N(\bq)$) controlling the population inversion on the external sphere can be written as $\sin(\theta_{N-1}(\bq)/2)$, which together with its partner $\phi_{N-1}(\bq)$ forms the exterior sphere such that $J[\phi_{N-1}(\bq),\sqrt{\frac{2(N-1)}{N}}S_{N^2-1}(\bq)+\frac{N-2}{N}]=J[\phi_{N-1}(\bq),\cos(\theta_{N-1}(\bq))]$. 

Furthermore, these $N-1$ nested two-spheres preserve the hierarchy among themselves. Explicitly, we start from a reference point $\bm{Q}$ with the wave function $\ket{\Psi_{\bm{Q}}}=\ket{1}=(0,0,\ldots0,1)^{\top}$ written in the local eigenstate basis, corresponding to the south pole of the exterior sphere $\bm{n}_{\rm ext}(\bm{Q})=-\bm{\hat{z}}$. In order to have a nontrivial Chern number, the wave function must become orthogonal to this reference point at least once within the BZ ($\bra{\Psi_{\bm{Q}}} \Psi_{\bm{Q}_{\perp1}} \ra=0$ for some $\bm{Q}_{\perp1}$), so that the exterior north pole is reached and $C$ is nonzero via Eq.(\ref{eq:WindingAroundSing_ChernNum}). Upon analyzing the $\hat{z}$-axes of the interior spheres, we find that a pole of the first interior sphere can be only covered if the exterior vector is inverted to the north pole ($\bm{n}_{\rm ext}(\bm{Q})=\bm{\hat{z}}$). In fact, as we move down to each layer, we see that one more pole-constraint is imposed to outer spheres~\cite{Supp}.
%both poles of the first interior sphere to be covered, there must be at least two distinct points in the BZ where the state vector gets inverted on the exterior sphere. In fact, as we move down to each inner layer, we see that one more condition is imposed at each stage. 

We numerically validate that the total Berry curvature separates into contributions coming from the nested spheres as in Eq.(\ref{eq:BerryCurv_Jacobians_anyN}) in the Harper-Hofstadter model with higher number of bands for $1/N$ flux-quantum per plaquette supporting a Chern insulating state with $C=1$ in its ground state. We observe that in general for any $N$ number of bands, a nontrivial Chern number is accompanied by a band inversion on the exterior sphere where the wave function visits both its poles and becomes orthogonal to itself similar to the two- and three-level cases.

 \subsection{Wilson Line Measurements}
Experimentally, the nested spheres configuration bears important consequences. For example, geometry of the Bloch states can be reconstructed through Wilson line measurements in ultracold quantum gases, as has been successfully implemented in a honeycomb optical lattice for $N=2$ by using  well-established band mapping techniques~\cite{Li16_Science_wilson}. In this experiment, an atomic cloud is prepared in the lower band at reference momentum $\bm{Q}$, where the initial wave function is given by $\ket{\Phi^1_{\bq}}=e^{i\bq.\hat{{\bm r}}} \ket{\Psi^1_{\bq}}$ with the superscript denoting the band index. Note that we here adopt the notation $\ket{\Psi^1_{\bq}}$ for the cell-periodic  part of the Bloch state $\ket{\Phi^1_{\bq}}$ to make the right correspondence with the previous sections. Since the Wilson line operator $\hat{\bm{W}}_{\bm{Q}\rightarrow\bm{q}}$ governs the dynamics in a degenerate subspace~\cite{WilczekZee84_PRL,AlexandradinataBernevig16_PRB,Alexandradinata14_PRB,BouhonSlager19_PRB}, the limit of effectively degenerate bands can be reached by accelerating the cloud to a final momentum $\bq$ under a suitably large force so that the band dispersion becomes negligible. Elements of the Wilson line matrix $\bra{\Phi^n_{\bq}} \hat{\bm{W}}_{\bm{Q}\rightarrow\bm{q}}\ket{\Phi^1_{\bq}} =\bra{\Psi^n_{\bq}}\Psi^1_{\bm Q}\ra$, which can be measured through band mapping, then give access to the cell-periodic part of the Bloch states since the Wilson line operator is essentially responsible for the quasimomentum change $\Delta\bq=\bq-\bm{Q}$ of the atoms. For the two-level case, this has allowed for mapping the geometry of the Bloch states across the BZ by comparing the eigenstates at all other quasimomenta $\bq$ with the reference states at $\bm{Q}$~\cite{Li16_Science_wilson}. 

The Wilson line tomography can be readily extended to the $N$-level case following our construction to reveal a nontrivial Chern number. When particles are suddenly moved between the two points in reciprocal space, measuring the overlap $|\bra{\Psi^1_{\bq}}\Psi^1_{\bm Q}\ra|=\sin(\theta_{\bq}/2)$ via band mapping directly gives access to the mixing angle on the exterior sphere. Since the exterior sphere plays a leading role in controlling the Chern number, one can then determine whether the wave function becomes orthogonal with respect to the reference state, where a vanishing $\theta_{\bq}$ would reveal information on nontrivial topology.

%%%%%%%%%%%%%%%%%%%%%%%%%%%%%%%%%%%%%%%%%%%%%%%%%%%%%%%%%%%%%%%%%%%%%
\section{Time Evolution and State Tomography}   \label{sec: tomography}
Although the hitherto unknown two-sphere structure for the Chern number in larger Hilbert spaces that we established in Secs.~\ref{sec: Nested_N=3} and \ref{sec: N-levels} presents a powerful way of characterizing the topological invariant in terms of effective band inversions primarily occurring on the exterior sphere, it concerns mainly the coherence vector components associated to the diagonal Gell-Mann matrices. The full coherence vector on the other hand carries all the information stored in the wave function and would allow access to the entire nested structure. We now turn our focus on the coherence vector defined on the generalized Bloch hypersphere, which provides a complete characterization of the quantum state also beyond the Chern number. Therefore, it would naturally be of high interest to extract the entire coherence vector in experiment. In the following, we demonstrate how the dynamics of a time-evolving state on the generalized Bloch hypersphere can be utilized for this purpose.

The time evolution of a quantum state with the density matrix given in Eq.(\ref{eq:rho def}) under the effect of the Hamiltonian (\ref{eq:Hamiltonian vec}) can be equally expressed in terms of generalized rotations of its coherence vector in $N^2-1$ dimensions upon invoking the von Neumann equation~\cite{Hioe81_PRL}, 
    \begin{equation}    \label{eq:vonNeumann}
        \frac{\partial\hat{\rho}}{\partial t}=i[\hat{\rho},\hat{H}]\implies\frac{\dd S_j}{\dd t}=2f^{jkl}h_{k}S_l \,. 
    \end{equation}
Here, $f^{jkl}$ is the anti-symmetric structure constants of the SU($N$) group, which for $N=2$ simply is the Levi-Civita symbol $\epsilon^{jkl}$, reducing Eq.(\ref{eq:vonNeumann}) to the familiar spinor precessing around the Hamiltonian vector $\bm{h}$ on the Bloch sphere. For $N\geq 3$, although still being totally anti-symmetric, $f^{jkl}$ is nonvanishing only for a subset of the non-repeating indices, whose value also differs along different directions (see Appendix \ref{Appx:GellManns _andRotations}). Consequently, under unitary dynamics the coherence vector explores only an asymmetric subspace of the hypersphere in $\mathds{R}^{N^2-1}$. It has been further shown~\cite{Hioe81_PRL} that there are $N$ constants of motion, given by $c_k=\tr[\hat{\rho}^k]$ for $1\leq k\leq N$, and the subspace to which the dynamics of the coherence vector are restricted has been studied exclusively for the $N=3$ case in Refs.~\cite{Khanna97_AnnalsPhys_GeometricPF,Kimura03_PLA}.

In the case of two levels, SU($2$) is the double cover of SO$(3)$ and the state vector can be conveniently expressed by using two angles on the Bloch sphere. This has been successfully employed in quantum simulation experiments, where these angles have been directly measured via state tomography techniques \cite{Alba11_PRL,Hauke14_PRL,Flaschner16_Sci,Li16_Science_wilson} by tracing the motion of the spinor precessing around the magnetic field as given in Eq.(\ref{eq:vonNeumann}). The ability of experimentally constructing the state vector (up to an overall phase) has allowed for probing critical topological phenomena that can not be revealed simply through band energy measurements~\cite{Cooper19_RMP,Tarnowski19_NatCom,Flaschner18_NatPhys_DynSingl,Unal19_PRR}. 

We here illustrate how the geometric insight obtained from the generalized Bloch hypersphere can be exploited to extend the state tomography to models featuring three bands. In an optical lattice with three sublattices labelled $(a,b,c)$ per unit cell, the momentum distribution $m(\bq)$  (defined as the square of the Fourier transform of the Bloch state) of a state $\ket{\Psi_{\bq}}$ can be measured via standard time-of-flight images~\cite{Cooper19_RMP}. Upon implementing the coherence vector description, we find that the momentum distribution in a lattice system corresponds to the purely real off-diagonal Gell-Mann matrices for any number of bands. For $N=3$, these are the first, fourth and the sixth components, which result in
\begin{eqnarray}  \label{eq:TOF n(q,t)}
        m(\bq)&=&\lvert\Tilde{w}(\bq)\rvert^2\lvert(\bra{a}+\bra{b}+\bra{c})\ket{\Psi_{\bq}}\rvert^2 \nonumber\\
        &=&\lvert\Tilde{w}(\bq)\rvert^2(1+S_1(\bq)+S_4(\bq)+S_6(\bq)) \,,
\end{eqnarray}
where $\Tilde{w}(\bq)$ is the Fourier transform of the Wannier centers~\cite{Flaschner16_Sci,Li16_Science_wilson,Hauke14_PRL,Cooper19_RMP}. For two-level systems, there is only one term ($S_1(\bq)=\bra{\Psi_{\bq}}\hat{\sigma}_x\ket{\Psi_{\bq}}$) in Eq.(\ref{eq:TOF n(q,t)}), while generally there will be $(N^2-N)/2$ components contributing to $m(\bq)$. 

Although Eq.(\ref{eq:TOF n(q,t)}) does not directly reveal individual $S_i(\bq)$ components, the full coherence vector can be reconstructed upon noticing that the dynamical rotations governed by Eq.(\ref{eq:vonNeumann}) couple only certain subspaces of the generalized Bloch hypersphere through nonvanishing $f^{jkl}$ terms. Namely, should the system be abruptly quenched to flat bands at time $t=0$, the Hamiltonian is diagonal and hence involves only the third and eighth Gell-Mann matrices, $\hat{H}(\bq)=h_3(\bq)\hat{\lambda}_3+h_8(\bq)\hat{\lambda}_8$. The coherence vector $\bm{S}(\bq,t)$ then starts evolving around these nonzero terms $h_3(\bq)$ and $h_8(\bq)$. The nonvanishing structure constants $f^{jkl}$ involving $3$ or $8$ as an index couple the components $1\leftrightarrow2$, $4\leftrightarrow5$ and $6\leftrightarrow7$ (see Appendix \ref{Appx:GellManns _andRotations}). Therefore, the time dependence (\ref{eq:vonNeumann}) of e.g.~the first and the second coherence vector components reads $\dot{S}_1(\bq,t)=2f^{123}h_3(\bq)S_2(\bq,t)$ and $\dot{S}_2(\bq,t)=2f^{231}h_3(\bq)S_1(\bq,t)$ with no $h_8(\bq)$ term since $f^{128}=0$, while similar differential equations govern the dynamics of $[{S}_4(\bq,t),{S}_5(\bq,t)]$ and $[{S}_6(\bq,t),{S}_7(\bq,t)]$. Solving Eq.(\ref{eq:vonNeumann}) for all these nonvanishing terms, we thus arrive at the simple oscillatory evolution for the coherence vector components relevant for the momentum distribution
\begin{eqnarray} \label{eq:osc.solutions of S1,4,6}
S_1(\bq,t)&=&\cos\left[2h_3(\bq)t\right]S_1(\bq,0)-\sin\left[2h_3(\bq)t\right]S_2(\bq,0), \nonumber \\
S_4(\bq,t)&=&\cos\left[(h_3(\bq)+\sqrt{3}h_8(\bq))t\right]S_4(\bq,0)  \nonumber \\
&& -\sin\left[(h_3(\bq)+\sqrt{3}h_8(\bq))t\right]S_5(\bq,0),  \nonumber \\
S_6(\bq,t)&=&\cos\left[(\sqrt{3}h_8(\bq)-h_3(\bq))t\right]S_6(\bq,0) \nonumber \\  &&-\sin\left[(\sqrt{3}h_8(\bq)-h_3(\bq))t\right]S_7(\bq,0),
\end{eqnarray}
which also results in oscillatory behavior in $m(\bq,t)$ at each quasimomentum $\bq$ upon inserting in Eq.(\ref{eq:TOF n(q,t)}). Most importantly, the generalized rotations of the coherence vector given in Eq.(\ref{eq:osc.solutions of S1,4,6}) is restricted to independent planes coupling only $(S_1,S_2),(S_4,S_5)$ and $(S_6,S_7)$ for quenches involving flat bands. 

We therefore devise a tomography protocol consisting of two stages which will allow us to resolve the full coherence vector at the initial time $t=0$ by monitoring the oscillatory time dependence of the momentum distribution Eqs.(\ref{eq:TOF n(q,t)}) and (\ref{eq:osc.solutions of S1,4,6}). Specifically,  %\newline
%\begin{itemize}

\textbf{Step 1:} Quench with \(h_3(\bq)=0\), where the Hamiltonian corresponds to flat bands with one of the sublattices kept at much lower energies [$H\propto\text{diag}(1,1,-2)$]. The resulting momentum distribution oscillates as
\begin{eqnarray*}  
        m(\bq,t) &\propto& 1+S_1(\bq,0) \nonumber \\ && +\cos\left[\sqrt{3}h_8(\bq)t\right] \left\{S_4(\bq,0)+S_6(\bq,0) \right\}   \nonumber \\
        && -\sin\left[\sqrt{3}h_8(\bq)t\right] \left\{S_5(\bq,0)+S_7(\bq,0) \right\} \,,
\end{eqnarray*}
from which one can gather \(S_1(\bq,0)\), \(S_4(\bq,0)+S_6(\bq,0)\) and \(S_5(\bq,0)+S_7(\bq,0)\);  %\newline

\textbf{Step 2:} Quench with \(h_3(\bq)=\sqrt{3}h_8(\bq)\), which corresponds to a flat band Hamiltonian [$H\propto\text{diag}(2,-1,-1)$] where two of the sublattices are held at much lower energy compared to the other one. Similarly, the time dependence of $m(\bq,t)$ reveals \(S_6(\bq,0)\), \(S_1(\bq,0)+S_4(\bq,0)\), and \(S_2(\bq,0)+S_5(\bq,0)\). 
%\end{itemize} 

Following these two quenches, one can then resolve the coherence vector components $S_1(\bq),S_4(\bq),S_6(\bq)$ independently at the initial time while also acquiring additional constraints on the rest. The remaining components can be found by either employing a third quench (with $h_3(\bq)=-\sqrt{3}h_8(\bq)$) or directly employing the below relations between the components,
\begin{eqnarray*}
S_8=-\frac{1}{2\sqrt{3}}&\Big(&1+3S_1 \pm\\
 && 3\sqrt{(1+S_1)^2-(S_4+S_6)^2-(S_5+S_7)^2}\Big) \,, 
\end{eqnarray*}
\begin{eqnarray*}
S_2&=&\frac{(S_4+S_6)(S_2+S_5)-S_4(S_5+S_7)}{S_4+S_6+\frac{2}{3}(1-\sqrt{3}S_8)}\,, \\
S_3&=&\frac{3(S_4^2+S_5^2-S_6^2-S_7^2)}{4(1-\sqrt{3}S_8)} \,,
\end{eqnarray*}
which necessarily hold for pure states~\cite{Supp}. We note that the sign ambiguity in the definition of $S_8$ does not pose a problem upon ensuring the continuity and smoothness following an initial sign choice, as well as the normalization of the coherence vector. Secondly, the Fourier transform of the Wannier centers enters as well to the momentum distribution as an overall envelop~\cite{Hauke14_PRL,Flaschner16_Sci,Cooper19_RMP}. This in principle can be also measured separately or determined through another quench.

We implement our tomography protocol in the Harper-Hofstadter Hamiltonian given in Eq.(\ref{eq:HofstHamiltonian}). Starting from the ground state at one-third filling, we quench to flat bands as proposed and let the system evolve for a tomography time $t$ to sample the oscillatory dependence of the momentum distribution following each quench as in a time-of-flight (TOF) measurement. By extracting the off-set, amplitude and phase shift of these oscillations, we numerically reconstruct the coherence vector at each momentum. We then calculate the Chern number of the ground state by using the numerically constructed $\bm{S}(\bq)$ vector through Eq.(\ref{eq:Chern_generalS}) and confirm the quantized value for the unit Chern number.

Our tomography scheme not only extends the protocols implemented in ultracold quantum gases~\cite{Alba11_PRL,Hauke14_PRL,Flaschner16_Sci,Li16_Science_wilson,Cooper19_RMP} to three-level systems, it also establishes the general structure of TOF measurements, as depicted in Eq.(\ref{eq:TOF n(q,t)}), for higher level systems and demonstrates how the dynamics restricted on sub-planes on the Bloch hypersphere can be employed for this purpose. General principles put to use through Eqs.(\ref{eq:vonNeumann},\ref{eq:osc.solutions of S1,4,6}) translate similarly to higher level systems although the generalized rotations would be more involved. In these settings, it could be particularly useful to exploit these relations in a clever way to extract only certain components of the coherence vector that could be related to the physics of interest. While existing tomography methods generally rely on measuring each element of the density operator to construct the full complex density matrix, our scheme targets directly the coherent dynamics of the real valued coherence vector. This requires monitoring time evolution of the state occurring in the shape of rotations and fitting a sinusoidal function to these oscillations.

%%%%%%%%%%%%%%%%%%%%%%%%%%%%%%%%%%%%%%%%%%%%%%%%%%%%%%%%%%%%%%%%%%%%
%%%%%%%%%%%%%%%%%%%%%%%%%%%%%%%%%%%%%%%%%%%%%%%%%%%%%%%%%%%%%%%%%%%%%
\section{Conclusion}   \label{sec: Conclusion}
We have employed a coherence vector description that is defined through the expectation values of the generalized Gell-Mann matrices of the SU$(N)$ algebra to characterize a quantum state in $N$-dimensional Hilbert space. While this picture is well-established for the case of two levels in terms of a three-dimensional vector on the familiar Bloch (two-)sphere, we have shown that its natural extension to $N$-level systems encompassing the coherence vector $\bm{S}$ living on a $(N^2-1)$-dimensional Bloch hypersphere provides unique insight. 
%The quantum metric and the Berry curvature take suggestive forms dictated by the underlying algebra.

Focusing on two spatial dimensions, we prove that the Chern number in  $N$-band models can be decoupled into contributions coming from $N-1$ two-spheres. Most importantly within our framework, the diagonal Gell-Mann matrices now take up a $\hat{z}$-axis role, similar to the part played by the Pauli matrix $\hat{\sigma}_z$ but indeed extending this character to higher Gell-Mann matrices. We show that among the nested two-spheres there exists an exterior sphere, on which the nontrivial invariant manifests itself through band inversions where the eigenvector in the multilevel system becomes orthogonal with respect to a reference point within the BZ. We have demonstrated that the Chern number can be fully expressed on the exterior sphere which dominates the topological characterization. We further consider the time evolution of the coherence vector in terms of generalized rotations on the Bloch hypersphere. By exploiting the fact that quenching to the flat bands results in rotations in certain subspaces in this higher dimensional plane, mixing only a subset of the coherence vector components, we develop a dynamical state tomography scheme to extract the full coherence vector in three-level systems. 

The exterior sphere for the Chern number in a multilevel system can be motivated by adiabatically (with respect to the lowest band) pushing away all higher bands except the first excited one. We imagine that the higher bands are allowed to transfer their topological charge to the lowest lying excited band before being pushed away as trivial. Throughout this process, the gap to the ground state has never been closed, therefore its Chern number $C$ remains the same even though the Berry curvature changes. We eventually arrive at the situation that all the topology manifests in the ground state ($C$) and the first excited ($-C$) bands, corresponding to the exterior sphere with the interior structures nullified. This procedure is robust and essentially demonstrates that a two-level model is sufficient to describe the Chern number in any Hilbert space dimensions. Formally, $C$ is defined as the U(1) phase that the wave function picks up upon circling the BZ, which is isomorphic to circle $\text{S}^1$ and in principle can be wrapped around the equator of a two-sphere (i.e.~the exterior sphere), effectively connecting to a magnetic monopole sitting at the center of the familiar Bloch sphere.
%finding incarnation as a magnetic monopole sitting at the center of the familiar Bloch sphere for two-band models. 
%We argue that this picture can be more general, where the U(1)$=\text{S}^1$ phase of the Chern number can be wrapped around the exterior two-sphere even in the many-band case, effectively connecting to a monopole.

Our work presents a geometric description for the topological classification in terms of the Chern number and the dynamics of multilevel systems. An interesting direction entails investigating hidden theoretical connections that such visual insight based on specific roles of the certain Gell-Mann matrices can unearth beyond the Chern number as well as new opportunities for experiments to probe the quantum geometry in any dimensions.

%TC:ignore
Data sharing not applicable to this article as no datasets were generated or analysed during the current study.

\begin{acknowledgments}
C.J.D.K.~and F.N.\"U.~thank Robert-Jan Slager for fruitful discussions and comments on the manuscript. F.N.\"U.~also acknowledges discussions with Andr\`e Eckardt, Georgia M.~Nixon and Ulrich Schneider. This work was supported by the Royal Society under a Newton International Fellowship, by EPSRC Grant EP/P009565/1 and by a Simons Investigator Award.

\end{acknowledgments}

\bibliography{references}

%%%%%%%%%%%%%%%%%%%%%%%%%%%%%%%%%%%%%%%%%%%%%%%%%%%%%
%\clearpage
\newpage
\appendix
\section{Gell-Mann matrices and SO$(N^2-1)$ rotations of the $N$-level Bloch vector} \label{Appx:GellManns _andRotations}
In this Appendix, we briefly summarize the structure of SU(3) as employed in the main text. The below eight traceless, Hermitian $3\times3$ operators form a trace-orthonormal basis of the Lie algebra \(\mathfrak{su}(3)\). General imaginary, linear combination of these vectors provides the axis of rotation, the magnitude of the superposition determines the angle of rotation and exponentiation then equates to the rotor which captures generic elements of the Lie Group SU(3). Gell-Mann matrices that we use in the main text are,
\begin{equation*}
    \begin{split}
        \hat{\lambda}_1=\begin{bmatrix}
0 & 1 & 0\\
1 & 0 & 0\\
0 & 0 & 0
\end{bmatrix}~~
 \hat{\lambda}_2=\begin{bmatrix}
0 & -i & 0\\
i & 0 & 0\\
0 & 0 & 0
\end{bmatrix}~~
 \hat{\lambda}_3=\begin{bmatrix}
1 & 0 & 0\\
0 & -1 & 0\\
0 & 0 & 0
\end{bmatrix}\\\
 \hat{\lambda}_4=\begin{bmatrix}
0 & 0 & 1\\
0 & 0 & 0\\
1 & 0 & 0
\end{bmatrix}~~
 \hat{\lambda}_5=\begin{bmatrix}
0 & 0 & -i\\
0 & 0 & 0\\
i & 0 & 0
\end{bmatrix}~~
 \hat{\lambda}_6=\begin{bmatrix}
0 & 0 & 0\\
0 & 0 & 1\\
0 & 1 & 0
\end{bmatrix}\\\
 \hat{\lambda}_7=\begin{bmatrix}
0 & 0 & 0\\
0 & 0 & -i\\
0 & i & 0
\end{bmatrix}~~
 \hat{\lambda}_8=\frac{1}{\sqrt{3}}\begin{bmatrix}
1 & 0 & 0\\
0 & 1 & 0\\
0 & 0 & -2
\end{bmatrix}.
    \end{split}
\end{equation*}
The trace-orthonormality condition of the group generators defines the structure of the bilinear form on the inner-product space: $\tr\{\hat{\lambda}_a\hat{\lambda}_b\}=2\delta_{ab}$.
%In this representation, it is matrix multiplication which defines the associative bilinear product (not to be confused with the bilinear form) of the algebra. 
The spanning property of the vector space basis (extended to include the $3\times3$ identity matrix) allows one to 
%express every output of the surjection \(\circ :V\times V\rightarrow V \) as a general linear combination over the complex scalars. Of course, the matrix product can be split into the symmetric anti-commutation part and the anti-symmetric commutation part.
split the matrix product into the symmetric anti-commutation part and the anti-symmetric commutation part as,
\begin{equation*}
    \begin{split}
       \hat{\lambda}_i\hat{\lambda}_j+\hat{\lambda}_j\hat{\lambda}_i=\{\hat{\lambda}_i,\hat{\lambda}_j\}=\frac{4}{3}\delta_{ij}\hat{\id}_3+2d^{ijk}\hat{\lambda}_k,\\\
       \hat{\lambda}_i\hat{\lambda}_j-\hat{\lambda}_j\hat{\lambda}_i=[\hat{\lambda}_i,\hat{\lambda}_j]=2if^{ijk}\hat{\lambda}_k.
        \end{split}
\end{equation*}
The Einstein summation convention is implied in the right-most side of both equations and likewise below,
\begin{equation*}
    \hat{\lambda}_i\hat{\lambda}_j=\frac{2}{3}\delta_{ij}\hat{\id}_3+d^{ijk}\hat{\lambda}_k+if^{ijk}\hat{\lambda}_k.
\end{equation*}
Although here given for SU(3), these structure constants can be similarly defined in any dimensions. In general, the symmetric structure constants $(d^{ijk})$ capture the degree to which the abstract vectors project onto one another with non-vanishing extent, whereas the anti-symmetric structure constants $(f^{ijk})$ capture the degree to which the abstract vectors reject one another. The anisotropy of the index structure characterises the lack of complete hyperspherical symmetry (for any $N>2$). For SU(3), nonzero components of the anti-symmetric structure constants are 
\begin{eqnarray*}
&&f^{123}=1; \\
&&f^{147}=f^{165}=f^{246}=f^{257}=f^{345}=f^{376}=1/2;\\
&&f^{458}=f^{678}=\sqrt{3}/2, 
\end{eqnarray*}
while the symmetric structure constants are 
\begin{eqnarray*}
&d^{118}&=d^{228}=d^{338}=-d^{888}=\frac{1}{\sqrt{3}}; \\
&d^{448}&=d^{558}=d^{668}=d^{778}=-\frac{1}{2\sqrt{3}}; \\ &d^{344}&=d^{355}=-d^{366}=-d^{377}=-d^{247}=d^{146}\\
&&=d^{157}=d^{256}=\frac{1}{2}. 
\end{eqnarray*}

The structure of the SU($N$) algebra dictates the quantum geometric tensor as given in the main text. Not only the Berry curvature but also the quantum metric inherits the geometry of the underlying algebra through the coherence vector~\cite{Graf21_arXiv_eigenprojectors}. Namely, the SU($N$) group features the additional structure constant which is totally symmetric in all three indices $d^{ijk}$ (which is essentially zero for two-level systems). Consequently, along with the usual Euclidean inner product $\bm{a}\cdot\bm{b}=a_{j}b_{j}$, and the skew-symmetric wedge-product $(\bm{a}\wedge \bm{b})_i=f^{ijk}a_{j}b_{k}$ defined through $f^{ijk}$ for any vectors $(\bm{a,b})$, SU($N$) algebra also harbors a symmetric vector star-product $(\bm{a}\star \bm{b})_i=d^{ijk}a_{j}b_{k}$ involving $d^{ijk}$. The symmetric products enter the real part of the QGT \cite{Graf21_arXiv_eigenprojectors}, forming the quantum metric via,
\begin{equation}
     g_{\alpha,\beta}(\bq)=\frac{N-1}{2N}  \partial_{\alpha} \bm{S}\cdot \partial_{\beta} \bm{S}-\frac{1}{4}\bm{S}\cdot\left(\partial_{\alpha}\bm{S}\star\partial_{\beta}\bm{S}\right)   \ \,.
\end{equation}
If the coherence vector is that of a pure state, the metric simplifies even further to involve only the dot product~\cite{Graf21_arXiv_eigenprojectors}.

In terms of the angle parameters introduced in Sec.\ref{sec: Nested_N=3}, the coherence vector defined on the generalized Bloch takes the form
\begin{equation}  
 \bm{S}=\left(
\begin{array}{ccc}
\cos^2(\theta/2)\sin(\vartheta)\cos(2\varphi) \\
\cos^2(\theta/2)\sin(\vartheta)\sin(2\varphi)\\
\cos^2(\theta/2)\cos(\vartheta) \\
\sin(\theta)\cos(\vartheta/2)\cos(\phi+\varphi) \\
\sin(\theta)\cos(\vartheta/2)\sin(\phi+\varphi) \\
\sin(\theta)\sin(\vartheta/2)\cos(\phi-\varphi) \\
\sin(\theta)\sin(\vartheta/2)\sin(\phi-\varphi) \\
(3\cos(\theta)-1)/(2\sqrt3)
\end{array}\right) .
\end{equation}

Turning to the time-evolution dynamics, unitary action on the Hilbert space for an $N$-level quantum system induces an \(SO(N^2-1)\) rotation on the corresponding coherence vector of the state as given in Eq.(\ref{eq:vonNeumann}). This can be seen more explicitly by writing Eq.(\ref{eq:vonNeumann}) in vectorial terms and juxtaposing the generator of the transformation with the coherence vector through left matrix action via,
\begin{equation*}
\frac{d\vec{S}}{dt}=\underline{R}\vec{S}\implies \vec{S}(t)={\cal T} e^{\int^{t}d\tau\underline{R(\tau)}}\vec{S}(0),
\end{equation*}
where ${\cal T}$ is time ordering and \([R]_{jl}=2f^{jkl}h_{k}\). For $N=2$, the $3\times3$ matrix $R$ is skew-symmetric, thus the matrix exponential is easily calculated using Rodrigues' formula. For $N=3$, the matrix is still skew-symmetric but the $8\times8$ nature of the matrix does not allow easy usage of Rodrigues' formula, however, it should be noted that there still exists a Rodrigues-like formula so long as the matrix obeys a set of conditions~\cite{Gallier}.

%\clearpage
%\newpage

\section{Representation of the wave function}  \label{Appx:Alternative Azimt.Angles}
We here discuss an alternative parametrization for a qutrit state involving four angles. 
As mentioned in the main text, the form given in Eq.(\ref{eq:Wf_3level_angles}) is not coincidental but in fact takes power from the general structure of the Gell-Mann matrices in our choice of basis. 
In this representation, the polar angle $\theta$ corresponds to mixing on the external sphere, whereas internal sphere angles are distributed symmetrically between the elements of the wave function, so that the Berry curvature decouples into contributions coming from an external and an internal two-sphere associated to diagonal Gell-Mann matrices as in Eq.(\ref{eq:BerryCurv_Jacobian_3level}). We now imagine a small difference where the internal sphere azimuthal angle $\varphi$ is introduced into the wave function on unequal footing,
%\begin{equation}
%    \ket{\Psi_{\bm q}}=\left(
%    \begin{array}{ccc}
%\cos(\theta_{\bm q}/2)\cos(\vartheta_{\bm q}/2)~e^{i(\phi_{\bm q}+\frac{1}{2}\varphi_{\bm q})}\\
%\cos(\theta_{\bm q}/2)\sin(\vartheta_{\bm q}/2)~e^{i(\phi_{\bm q}-\frac{1}{2}\varphi_{\bm q})}\\
%\sin(\theta_{\bm q}/2)
%\end{array}\right) \end{equation}
%with the angles defined in the range \(0\leq[\theta,\vartheta]<\pi\), \(0\leq\phi<2\pi\) and \(0\leq\varphi<4\pi\).
%The Berry curvature is given by
%\begin{equation}
%\begin{split}
%    \Omega=\frac{1}{2}J\left[\phi,\frac{2}{\sqrt{3}}S_8+\frac{1}{3}\right]+J\left[\frac{1}{2}\varphi,S_3\right]
%\end{split} \end{equation}
%That is, \(\frac{1}{2}\varphi\) functions as the internal azimuthal angle but now the Chern number of the corresponding internal 2-sphere is twice the number of wrappings. Now if we apply the local unitary transformation
%\begin{equation}
%    U_q=\exp\left[\frac{i}{2}\varphi_q\lambda_3\right]
%\end{equation}
%The local Berry curvature transforms as
%\begin{equation}
%    \Omega_q'=\Omega_q+\hat{z}\cdot\left[\nabla_q\times\bra{\Psi_q}iU_q^{\dagger}(\nabla_q U_q)\ket{\Psi_q}\right]
%\end{equation}
%The vortex term is cancelled if the azimuthal angle's range is halved and the Berry curvature reduces to (5). However, if we apply another local unitary transformation, namely
%\begin{equation}
 %   U_q=\exp\left[\frac{i}{2}\varphi_q\id_2\right]
%\end{equation}
%and again halve the azimuthal angle range then the state vector and Berry curvature become

\begin{equation}  \label{appx.eq._PsiAlternative}
    \ket{\Psi'_{\bm q}}=\left(\begin{array}{ccc}
\cos(\theta_{\bm q}/2)\cos(\vartheta_{\bm q}/2)~e^{i(\phi_{\bm q}+\varphi_{\bm q})}\\
\cos(\theta_{\bm q}/2)\sin(\vartheta_{\bm q}/2)~e^{i\phi_{\bm q}}\\
\sin(\theta_{\bm q}/2)\end{array}\right)
\end{equation}
where the angles span the same range \(0\leq[\theta,\vartheta]\leq\pi\) and \(0\leq[\phi,\varphi]<2\pi\) with Eq.(\ref{eq:Wf_3level_angles}). Note that this form is still general and any normalized quantum state can be defined in this way. The Berry curvature of this state $\ket{\Psi'_{\bm q}}$ can be calculated as,
\begin{equation}  
  \Omega'_{\bm q}=\frac{1}{2}J\left[\phi_{\bm q},\frac{2}{\sqrt{3}}S_{8}+\frac{1}{3}\right] +\frac{1}{2}J\left[\varphi_{\bm q},\cos(\Theta(\theta_{\bm q},\vartheta_{\bm q}))\right],
\end{equation}
for
\begin{equation}
\begin{split}
    \cos(\Theta(\theta_{\bm q},\vartheta_{\bm q}))=2\cos^2(\theta_{\bm q}/2)\cos^2(\vartheta_{\bm q}/2)-1,\\\
    \frac{2}{\sqrt{3}}S_{8}+\frac{1}{3}=\cos\theta_{\bm q}.
\end{split}
\end{equation}
While now the internal sphere $\hat{z}$-axis does not correspond to $S_3(\bq)$ and the third Gell-Mann matrix given in Appendix \ref{Appx:GellManns _andRotations}, the external sphere structure remains the same.

In our formulation given in the main text, the external sphere vector $\bm{n}_{\rm ext}$ started off at its south pole and the internal vector $\bm{n}_{\rm int}$ began on its equator. In the alternative formulation in Eq.B2, both these vectors begin at their south pole where the azimuthal angle is intrinsically not well-defined and we see that both vectors now have this commensurate starting point. The constraint of inverting the state on the external sphere for nontrivial Chern number is the same as before (i.e. $\theta_{\bm{Q}_{\perp}}=0$ for some quasi-momentum in the BZ). The internal sphere north pole is reached when
$\theta_{\bm{Q}_{\perp i}}=\vartheta_{\bm{Q}_{\perp i}}=0$ for any $\bm{Q}_{\perp i}$, which still requires the external vector to be on its north pole $\bm{n}_{\rm ext}=+\hat{z}$, hence a second order condition. 
%The constraint on the internal sphere is now that only \(\theta(q_c)=\vartheta(q_c)=0\) (i.e. \(\lvert A_1(q_c)\rvert=1\)) whereas before it was both \(\theta(q_{c1})=\vartheta(q_{c1})=0\) (i.e. \(\lvert A_1(q_{c1})\rvert=1\)) and \(\theta(q_{c2})=0\), \(\vartheta(q_{c2})=\pi\) (i.e. \(\lvert A_2(q_{c2})\rvert=1\)) and this goes about symmetrising the number of conditions on each sphere, the external sphere still has a single first order constraint but the internal sphere also only has a single constraint albeit of second order. 

We can easily see an algorithm for extending this embedding to general $N$-level normalised Hilbert Space vectors via induction (for a given momentum state) such as,
\begin{equation}
    \ket{\Psi'_N}=e^{i\phi_{N-1}}\cos(\theta_{N-1}/2)\ket{\Psi'_{N-1}}\oplus\sin(\theta_{N-1}/2)\ket{1}
\end{equation}
for \(N\geq2\), and we define $\ket{\Psi'_1}=\ket{1}=[1]$, $\theta_0=\pi$.

%Finally, the double qubit decomposition now changes from (C5) to
%\begin{equation}
%    \begin{split}
%        \ket{\Psi_{Global}}=\left[\begin{array}{cc}
%         \cos(\theta/2)~e^{i\phi} \\
%             \sin(\theta/2) 
%        \end{array}\right]\\\
%         \ket{\Psi_{Internal}}=\left[\begin{array}{cc}
%         \cos(\theta/2)\cos(\vartheta/2)~e^{i\varphi} \\
%\sqrt{1-\cos^2(\theta/2)\cos^2(\vartheta/2)}    %   \end{array}\right]
%    \end{split} \end{equation}

\begin{widetext}
\section{Wave function representation for N=4 and N=5 levels}  \label{Appx: WaveFunction N=4,5}
Extending the description given in Eq.(\ref{eq:Wf_3level_angles}) for a three-level wave function, the four-level state vector can be parametrized as,
\begin{equation} \label{eq.4-level_W.f._angles}
    \ket{\Psi}=\left(\begin{array}{cccc}
    \cos(\theta_3/2)\cos(\theta_2/2)\cos(\theta_1/2)~e^{i(\phi_3+\phi_2+\phi_1)}  \\
    \cos(\theta_3/2)\cos(\theta_2/2)\sin(\theta_1/2)~e^{i(\phi_3+\phi_2-\phi_1)}\\
    \cos(\theta_3/2)\sin(\theta_2/2)~e^{i(\phi_3-2\phi_2)}\\
     \sin(\theta_3/2)\end{array}\right) .
\end{equation}
Similarly, the five-level state vector can be written as,
\begin{equation}
    \ket{\Psi}=\left(\begin{array}{ccccc}
    \cos(\theta_4/2)\cos(\theta_3/2)\cos(\theta_2/2)\cos(\theta_1/2)~e^{i(\phi_4+\phi_3+\phi_2+\phi_1)}  \\
    \cos(\theta_4/2)\cos(\theta_3/2)\cos(\theta_2/2)\sin(\theta_1/2)~e^{i(\phi_4+\phi_3+\phi_2-\phi_1)}  \\
    \cos(\theta_4/2)\cos(\theta_3/2)\sin(\theta_2/2)~e^{i(\phi_4+\phi_3-2\phi_2)}\\
    \cos(\theta_4/2)\sin(\theta_3/2)~e^{i(\phi_4-3\phi_3)}\\
     \sin(\theta_4/2)\end{array}\right).
\end{equation}
We now focus on the four-level wave function and analyse it in more detail, while noting that the $\hat{z}$-axes of the nested spheres and the hierarchy among them for the five-level case can be found in the similar way.

The corresponding coherence vector for the four-level state (\ref{eq.4-level_W.f._angles}) is given by,
\begin{equation}  
 \bm{S}=\left(
\begin{array}{ccc}
\cos^2(\theta_3/2)\cos^2(\theta_2/2)\sin(\theta_1)\cos\left[2(\phi_3+\phi_2)\right] \\
\cos^2(\theta_3/2)\cos^2(\theta_2/2)\sin(\theta_1)\sin\left[2(\phi_3+\phi_2)\right]\\
\cos^2(\theta_3/2)\cos^2(\theta_2/2)\cos^2(\theta_1) \\
\cos^2(\theta_3/2)\sin(\theta_2)\cos(\theta_1/2)\cos(\phi_3-\phi_2+\phi_1) \\
\cos^2(\theta_3/2)\sin(\theta_2)\cos(\theta_1/2)\sin(\phi_3-\phi_2+\phi_1) \\
\cos^2(\theta_3/2)\sin(\theta_2)\sin(\theta_1/2)\cos(2\phi_3-\phi_2-\phi_1) \\
\cos^2(\theta_3/2)\sin(\theta_2)\sin(\theta_1/2)\sin(2\phi_3-\phi_2-\phi_1)  \\
(\cos^2(\theta_3/2)\cos^2(\theta_2/2)-2\cos^2(\theta_3/2)\sin^2(\theta_2/2))/\sqrt3 \\
\sin(\theta_3)\cos(\theta_2)\cos(\theta_1/2)\cos(\phi_3+\phi_2+\phi_1) \\
\sin(\theta_3)\cos(\theta_2)\cos(\theta_1/2)\sin(\phi_3+\phi_2+\phi_1) \\
\sin(\theta_3)\cos(\theta_2)\sin(\theta_1/2)\cos(\phi_3+\phi_2-\phi_1) \\
\sin(\theta_3)\cos(\theta_2)\sin(\theta_1/2)\sin(\phi_3+\phi_2-\phi_1) \\
\sin(\theta_3)\sin(\theta_2/2)\cos(\phi_3-2\phi_2) \\
\sin(\theta_3)\sin(\theta_2/2)\sin(\phi_3-2\phi_2) \\
(1-4\sin^2(\theta_3/2))/\sqrt{6}
\end{array}\right) .
\end{equation}

The  four-level Berry curvature as given in Eq.(\ref{eq:BerryCurv_Jacobians_anyN}) is decomposed into contributions coming from three two-spheres; the external, internal and innermost spheres, whose $\hat{z}$-axes are given by,
\begin{equation}
\begin{split}
     z_{\rm ext}=\sqrt{\frac{3}{2}}S_{15}+\frac{1}{2}=\cos\theta_3\,,\\\
     z_{\rm int}=\frac{2\cos^2(\theta_3/2)}{3}\left(\cos^2(\theta_2/2)-2\sin^2(\theta_2/2)\right)+\frac{1}{3}\,,\\\
     z_{\rm innermost}=S_3=\cos^2(\theta_3/2)\cos^2(\theta_2/2)\cos\theta_1 \,.
\end{split}
\end{equation}

  \begin{figure*}
	\includegraphics[width=1\linewidth]{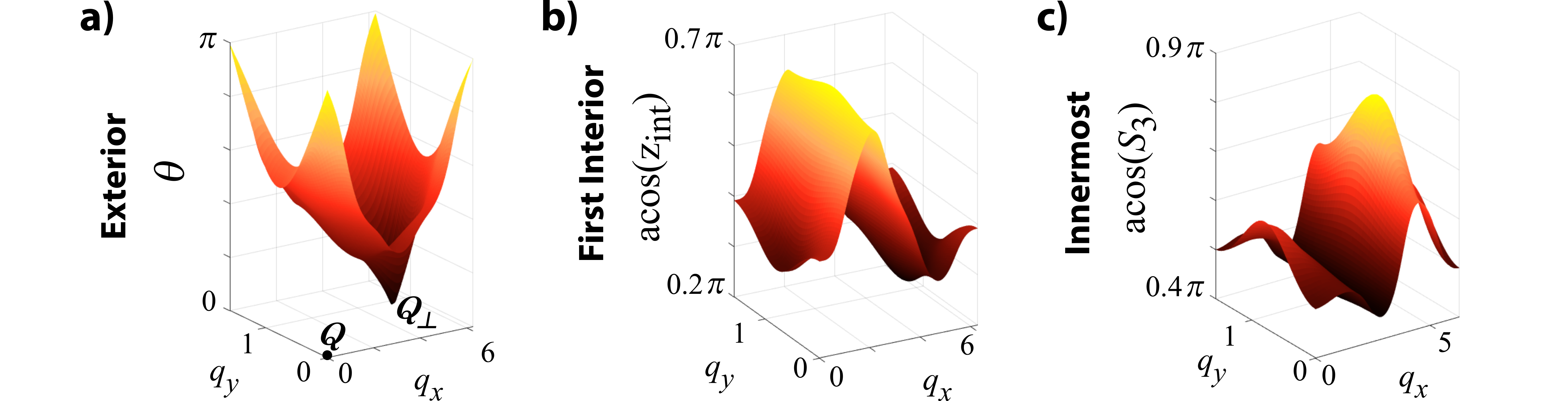}
	\caption{ Distribution of the polar angles on the nested spheres for the ground state of the Harper-Hofstadter model with $1/4$-flux quantum per plaquette, with reference to the eigenstates at ${\bm Q}=\Gamma$, as a function of crystal momentum $\bq$ with lattice constant set to one similar to the figures in the main text. $\hat{z}$-axes of the nested spheres are defined through Eq.(\ref{eq.4-level_W.f._angles}). (a) Starting from the exterior south pole ($\theta_{\bm Q}=\pi$), the state vector reaches the exterior north pole ($\theta_{\bm{Q}_{\perp}}=0$) once within the BZ, a prerequisite for nontrivial $C$, (b,c) while the interior sphere poles are not covered. Similarly, for the ground state ($C=1$) of the Harper-Hofstadter model with larger number of bands ($N\geq4$), we generically observe that the exterior sphere vector gets inverted to its north pole once, while the interior sphere poles are not reached.}
	\label{fig_appx_Hofst4}
\end{figure*}

When written in the eigenstate basis at some reference quasimomentum $\bm{Q}$, we start from the south pole of the external sphere, $\theta_3(\bm{Q})=\pi$. While reaching the external north pole again requires only $\theta_3(\bm{Q}_{\perp i})=0$, poles of the first interior sphere depends on a second order condition. The maximum and minimum values of $z_{\rm int}$ are reached only if $\theta_3(\bm{Q}_{\perp j})=0$ while simultaneously $\theta_2(\bm{Q}_{\perp j})=0$ (north pole) and $\theta_2(\bm{Q}_{\perp j})=\pi$ (south pole). When we move one more level down, poles of the innermost sphere impose a third order constraint. First, both spheres outer to it must be at their north poles $\theta_3(\bm{Q}_{\perp j})=\theta_2(\bm{Q}_{\perp j})=0$; only after which $\theta_1(\bm{Q}_{\perp j})=0$ and $\theta_1(\bm{Q}_{\perp j})=\pi$ correspond to the north and south poles of the innermost sphere. Hence the nested hierarchy among the two-spheres is still present also for the four-level case, which renders reaching the poles of the interior spheres less likely at each stage moving down the nested structure as one more constraint is introduced at each level. We demonstrate the distribution on these nested spheres for the ground state of the Harper-Hofstadter model with $1/4$-flux quantum per plaquette in Fig.~\ref{fig_appx_Hofst4}, which supports a topologically nontrivial state with $C=1$.

Just as coordinate singularities naturally arise when using one chart from the atlas to describe the entire two-sphere, 
%even if the open subset is the entire manifold, 
so too does our one-chart representation admit coordinate singularities at the hierarchical `poles'. When \(\theta_3=\pi\); \(~\theta_2\) and \(\theta_1\) (as well as all of the azimuthal angles) possess ill-defined value. When \(\theta_3=0\) it is only \(\phi_3\) that necessarily suffers from ill-definedness. 
When \(\theta_2=\pi\); 
\(~\theta_1\), (\(\phi_3+\phi_2+\phi_1\)) and (\(\phi_3+\phi_2-\phi_1\)) are all ill-defined but \(\theta_3\) (necessarily) and (\(\phi_3-2\phi_2\)) (potentially) are well-defined, when \(\theta_2=0\) it is only (\(\phi_3-2\phi_2\)) that is ill-defined. The form (\ref{appx.eq._PsiAlternative}) admits greater clarity for the extension of this property to the general case; for an N-level quantum system, \(\theta_j=\pi\) induces ill-definedness on all \(\theta_{k<j}\) and all \(\phi_{m\leq j}\) but \(\theta_j=0\) induces a (potentially) well-defined \(\phi_{j+1}+\phi_j\) and all other angles are (potentially) well-defined...aside from when \(j=N-1\) in which case \(\theta_{N-1}=0\) only induces an ill-defined \(\phi_{N-1}\).

\end{widetext}

\section{Some relations concerning the coherence vector} \label{Appx:S_relations}
Expressing our pure state vector as a normalised column vector, then computing the coherence vector through the expectation value of the Gell-Mann opertor array gives us:
    \begin{equation}
    \label{eq:Cohere}
        \ket{\Psi}=\left( \begin{array}{ccc}
\lvert A\rvert e^{i\theta_1} \\
\lvert B\rvert e^{i\theta_2} \\
\lvert C\rvert e^{i\theta_3} \end{array} \right) \implies\vec{S}=\left( \begin{array}{cccccccc}
2\lvert AB\rvert\cos(\theta_1-\theta_2) \\
2\lvert AB\rvert\sin(\theta_1-\theta_2) \\
\lvert A\rvert^2-\lvert B\rvert^2\\
2\lvert AC\rvert\cos(\theta_1-\theta_3)\\
2\lvert AC\rvert\sin(\theta_1-\theta_3)\\
2\lvert BC\rvert\cos(\theta_2-\theta_3)\\
2\lvert BC\rvert\sin(\theta_2-\theta_3)\\
\frac{1}{\sqrt{3}}\left(1-3\lvert C\rvert^2\right)
\end{array} \right)
    \end{equation}
where,
\begin{equation*}
    \lvert A\rvert^2+\lvert B\rvert^2+\lvert C\rvert^2=1
\end{equation*}
The form of Eq.(\ref{eq:Cohere}) allows one to easily derive the below relations that we employed in the main text:
\begin{equation*}
    \begin{split}
        S_1^2+S_2^2+S_3^2=\left(\lvert A\rvert^2+\lvert B\rvert^2\right)^2=\left(\frac{2}{3}+\frac{\sqrt{3}}{3}S_8\right)^2\\\
        S_4S_6+S_5S_7=4\lvert C^2 AB\rvert\cos(\theta_1-\theta_2)=\frac{2S_1}{3}(1-\sqrt{3}S_8)
    \end{split}
\end{equation*}
For three-level systems, the norm of the coherence vector is fixed to be
\begin{equation*}
    \sum_{i=1}^{8}S_i^2=\frac{4}{3}.
\end{equation*}
This can be rewritten as
\begin{equation*}
    \begin{split}
        \left(\frac{2}{3}+\frac{\sqrt{3}}{3}S_8\right)^2+(S_4+S_6)^2+(S_5+S_7)^2+S_8^2\\\ =\frac{4}{3}+2(S_4S_6+S_5S_7),
    \end{split}
\end{equation*}
where the last term can be also replaced by the expression in terms of \(S_1\) and \(S_8\). This quadratic equation can be solved to find \(S_8\) in terms of \(S_1\), \(S_4+S_6\) and \(S_5+S_7\) as given in Sec. \ref{sec: tomography} of the main text. 

Another set of relations can be found as
\begin{equation*}
    \begin{split}
        \frac{2}{3}(1-\sqrt{3}S_8)S_2=S_5S_6-S_4S_7 ,
    \end{split}
\end{equation*}
where we can trivially add equivalent terms to the left-hand side and right-hand side as
\begin{equation*}
    \begin{split}
        S_2S_4+S_2S_6+\frac{2}{3}(1-\sqrt{3}S_8)S_2=\\\ S_5S_6-S_4S_7+S_2S_4+S_2S_6+S_4S_5-S_4S_5=\\\
        (S_4+S_6)(S_2+S_5)-S_4(S_5+S_7).
    \end{split}
\end{equation*}
This then gives us a relation of \(S_2\) in terms of \(S_4+S_6\), \(S_2+S_5\), \(S_5+S_7\) and \(S_8\) as given in Sec. \ref{sec: tomography}.

The third and final relation utilized for the state tomography is
\begin{equation*}
    \begin{split}
        S_4^2+S_5^2-(S_6^2+S_7^2)=4\lvert C\rvert^2(\lvert A\rvert^2-\lvert B\rvert^2)=\\\ ...=\frac{4S_3}{3}(1-\sqrt{3}S_8),
    \end{split}
\end{equation*}
which gives \(S_3\) in terms of the $4^{th},5^{th},6^{th},7^{th}$ and $8^{th}$ coherence vector components.

Furthermore, it can be noted that the entire coherence vector and the wave function can be specified by simply using the $4^{th},5^{th},6^{th},7^{th}$ and $8^{th}$ components, which can be then re-expressed as,
    \begin{equation*}
        \ket{\Psi}=\frac{\sqrt{3}}{2\sqrt{1-\sqrt{3}S_8}}\left( \begin{array}{ccc}
S_4+iS_5 \\
S_6+iS_7 \\
\frac{2}{3}(1-\sqrt{3}S_8) \end{array} \right) .
    \end{equation*}

%TC:endignore
\end{document}